\newcommand{\beq}{\begin{equation}}
\newcommand{\eeq}{\end{equation}}

\newcommand{\bear}{\begin{eqnarray}}
\newcommand{\eear}{\end{eqnarray}}

\newcommand{\tn}{\textnormal}

\documentclass[review]{elsart5p}

\usepackage{graphicx}

\usepackage{amssymb}
\usepackage{bbding}
\usepackage{lineno}

\begin{document}

\begin{frontmatter}



\title{Signal coupling and signal integrity in multi-strip Resistive Plate Chambers
used for timing applications}

\author[1,2]{Diego Gonzalez-Diaz \corauthref{aut1}},
\corauth[aut1]{Corresponding author.}
\ead{D.Gonzalez-Diaz@gsi.de}
\author[2]{Huangshan Chen},
\author[2]{Yi Wang}
\address[1]{GSI Helmholtzcenter for Heavy Ion Research, Darmstadt, Germany}
\address[2]{Department of Engineering Physics, Tsinghua University, Beijing, China}

\begin{abstract}

We have systematically studied the transmission of electrical signals along 
several 2-strip Resistive Plate Chambers (RPCs)
in the frequency range $f=0.1-3.5$GHz. Such a range was chosen to fully cover the bandwidth associated
to the very short rise-times of signals originated in RPCs used for sub-100ps timing applications. This work
conveys experimental evidence of the dominant role of modal dispersion in counters built 
at the 1 meter scale, a fact that results in large cross-talk levels and strong signal
shaping. It is shown that modal dispersion appears in RPCs
due to the intrinsic unbalance between the capacitive and the inductive coupling
$C_m/C_o \neq L_m/L_o$.
A practical way to restore this symmetry
has been introduced (hereafter `electrostatic compensation'), allowing
for a cross-talk suppression factor of around $\times 12$ and a rise-time reduction
by 200ps. Under conditions of compensation the signal transmission is only limited by dielectric losses, yielding a 
length-dependent cutoff frequency of around 1GHz per 2 meter for typical float glass -based RPCs
($\tan \delta|_{glass} = 0.025\pm0.005$). 

It is further shown that
 `electrostatic compensation' can be achieved for an arbitrary number of strips as
long as the nature of the coupling is `short-range', that is an almost exact assumption
for typical strip-line RPCs. Evidence for deviations from the dominant TEM propagation 
mode has been observed, although they seem to
have negligible influence in practical signal observables. This work extends
the bandwidth of previous studies by a factor of almost $\times 20$.

\end{abstract}

\end{frontmatter}


\section{Introduction}

Cross-talk and signal integrity are, a priori, critical aspects for the operation of
electrically-long Resistive Plate Chambers (RPCs) with readout based on long multi-segmented electrodes.
The sometimes called multi-strip multi-gap RPCs (MMRPCs in short), pioneered by the
4$\pi$-experiment \cite{FOPI} at GSI fall, exemplarily, into this (by no means small) category.
In order to streamline the forthcoming discussion some preliminary considerations
are needed. We need first to introduce, as in \cite{Diego,Clayton},
the electrical length $\Lambda_e$ of a Resistive Plate Chamber.
This is done based on the `cutoff frequency' of the current induced by a single-electron
avalanche in the absence of Space-Charge effects and at typical working
conditions:
\bear
&&I(t)=e^{St} ~ \theta(t)  \label{exp1} \\
&&|\tn{ft}(I(t))| = \frac{1}{\sqrt{(2\pi f)^2 + S^2}} \label{frec-resp}
\eear
where $\theta(t)$ represents the Heavyside step function,
`ft' denotes the Fourier Transform, $||$ refers to the modulus of the bracketed
complex function, and $S=(\alpha-\eta)v_{drift}$ stands for the ionization rate in the active
gas. The latter serves as a definition, in the present context, of $\alpha$ and $\eta$
(the multiplication and attachment coefficients, respectively) and $v_{drift}$
(the drift velocity of the electron swarm). Curiously, since no assumption is made on the sign of $S$, 
eq. \ref{frec-resp} is obviously
identical to the modulus of the response function of a low-pass RC-circuit with $S=-1/RC$. 

The
characteristic `cutoff frequency' $f_c$ is usually defined as the frequency needed for a drop by
a factor of $1/\sqrt{2}$ (i.e., -3dB) with respect to the DC/zero-frequency limited
in the Fourier amplitude spectra, also
often equivalently as the frequency needed for a factor $1/2$ drop
in the Fourier power spectra, yielding thus $f_c=S/2\pi$ from eq. \ref{frec-resp}. Assuming, with sufficient
generality for present purposes, that signal propagation takes place at approximately half the speed of light, the electrical
length of an RPC of length $D$ can be estimated as:
\beq
\Lambda_e = \frac{D}{\lambda_c} = \frac{f_c}{v}D \simeq \frac{S}{\pi c}D \label{elec_length}
\eeq
where $c$ is the
speed of light and $D$
is the counter length. 

With approximate character, an electrically-long structure is customary defined as that fulfilling the condition
$\Lambda_e\!>\!1$ \cite{Clayton}. It is indeed possible to quantitatively understand the implication
of the above fact thanks to the recently
measured parameters of the swarm \cite{Urquijo} for the main RPC gas component (C$_2$H$_2$F$_4$). These measurements
were performed at low-pressure and are thus largely free of saturation effects, corresponding approximately to
the situation sketched in eq. \ref{exp1}. When extrapolated to typical working conditions ($T=20$$^\circ$C, $P=1$atm),
the parameters measured in \cite{Urquijo} provide the following lower bounds for the length $D^*$ above which an RPC
can be considered to be electrically-long:

\bear
&&D_{trigger}^*(E\!=\!50\tn{kV/cm})\!=\!\frac{\pi c}{S(E\!=\!50\tn{kV/cm})}  \! =  80\tn{cm} \label{Dstar1}\\
&&D_{timing}^*(E\!=\!100\tn{kV/cm})\!=\!\frac{\pi c}{S(E\!=\!100\tn{kV/cm})} \! =  5\tn{cm}  \label{Dstar2}
\eear
Eqs. \ref{Dstar1}, \ref{Dstar2} have been evaluated for two typical field values corresponding to
RPCs used for trigger, $E\simeq 50$kV/cm \cite{Santonico}, and timing, $E\simeq 100$kV/cm \cite{Fonte}. 
In particular, the cutoff frequency obtained for timing RPCs is as high as $f_c\simeq3$GHz.
The signal bandwidth is thus reasonably
 met by the fast front end electronics used in the so far existing strip timing RPC walls from
 the $4\pi$-experiment\cite{FOPI_FEE} (BW$\simeq1.5$GHz) and HADES\cite{HADES_FEE} (BW$\simeq2$GHz).
More realistically, rise-time measurements
performed directly over RPC signals with a 2-stage $\sim\!2$GHz-bandwidth amplifier at typical pressure, gas mixture and HV,
as in \cite{Paulo_IEEE}, have yielded a value of $f_c=1.4$ GHz at 100kV/cm. The 1/2 lower signal
bandwidth as compared to the one from \cite{Urquijo} may be interpreted as a contribution of both
the presence of Space-Charge for typical working
thresholds and the limited bandwidth of the readout system,
as has been argued by authors \cite{Lippmann}.\footnote{The estimate of $f_c$ from the signal rise-time
is not affected by a mere bandwidth reduction of the measuring device as long as the signal is purely exponential. 
This has been shown in \cite{Paulo_IEEE} for lineal devices. Deviations from the 
exponential growth at the discrimination point are thus needed in order to reduce $f_c$.}

It is thus expected, on very general grounds, that extreme conditions for signal transmission
take place, most prominently, in the 0.9m-long multi-strip wall of the 4$\pi$-experiment
\cite{FOPI} ($\Lambda_e =18$), the 1.6m-long 2-strip counter developed by Fonte in
2002 \cite{Fonte_long} and, more recently, in the 1.6m-long counters of
the EEE project \cite{EEE}(both $\Lambda_e =32$).\footnote{The HADES wall is not included in
this list due to its single-strip design, that
makes signal transmission a far simpler problem. HADES longest strip is $\Lambda_e \simeq 11$-long.}
Second generation multi-strip counters
like CBM-ToF \cite{CBMingo}, R3B-neuLAND \cite{R3B_neuLand},
R3B-iTOF \cite{R3B_itof} or the STAR-MTD upgrade \cite{STAR_MTD} are presently planned to be
built based on multi-strip designs with lengths in the range 0.5-2m ($\Lambda_e =10$-$40$). Although
single-strip designs like the HADES time-of-flight (ToF) wall offer a safe alternative from the point of view of signal
transmission, being virtually cross-talk free \cite{Blanco}, it is
at present questionable whether such an approach is practical for large numbers
of cells, in view of the `manpower$\times$time' overhead.

Of the above experiments, intrinsically multi-hit environments like CBM (3-5\% occupancy per strip per event,
$\simeq 1000$ tracks per event), speak against large cluster sizes that may typically arise either during
signal induction or due to sustained capacitive and inductive coupling during signal transmission
over many electrical lengths, i.e., cross-talk.
Efficiency measurements on the 2-strip, 1.6m-long prototype
of \cite{Fonte_long} point, indeed, to very high cross-talk levels (80-90\%)
while the $4 \pi$-experiment has recently reported a cluster size of the order of 4.5 strips/track \cite{Mladen}.
There is additional evidence that the cluster sizes measured in the earlier case are generated indeed during signal
transmission \cite{Diego} while the ones measured in the latter have
certainly a sizable contribution from the avalanche footprint (see next section).
Remarkably, it has been shown by the EEE collaboration that reducing the
system bandwidth from $f_c\simeq 3$GHz to $f_c\simeq300$MHz
(amplifier peaking time $\simeq 1$ns), i.e.,
 a $\sim 1/10$ bandwidth reduction, is compatible
with preserving a good time resolution and efficiency for long strips, thus
largely relaxing condition (\ref{Dstar2}) by approximately the same factor and easing
transmission.\footnote{As recalled in an earlier footnote, this statement is dependent, in detail, 
on the signal shape.}
Yet, the low filling factor (strip to pitch ratio) of
 less than 80\% used in those counters \cite{EEE} (presumably stemming from cross-talk optimization) together with the observed position-dependent space-resolution along the strips do not ensure the
uniformity of response; the absence of experimental information on the shear cross-talk levels and
cluster sizes do not allow to take a good multi-hit performance for granted, either.
Thus, none of these scenarios seem to be technologically satisfactory for a high multiplicity experiment like, for
instance, CBM.

Besides the aforementioned developments, systematic studies on signal propagation in RPCs are very limited. Numerical
simulations have been performed for the Pestov counter \cite{Pest_counter}
while systematic measurements are available for trigger-type RPCs \cite{ammosov}, where the main
phenomena ruling signal coupling in multi-strip counters have been unambiguously identified.
The work of W. Riegler in 2002 \cite{Riegler_transm}, combining experimental observations
and simulations, remains possibly the most complete up to date, despite covering a modest
200MHz bandwidth and being performed for trigger-type RPCs.
Not only cross-talk but specially dielectric losses (due to the presence of float glass) remain thus to be
assessed in timing-type RPCs up to $f_c \simeq 3$GHz.

This work is structured as follows: in section \ref{section1} we discuss cluster sizes in multi-strip RPCs,
identifying the contributions from
the avalanche induction profile (charge sharing)
and from the signal transmission (cross-talk). In section \ref{section2}
we present systematic pulser/scope-based measurements of transmission properties for 2-strip RPCs;
based on the structure of the solutions for the loss-less transmission line problem \cite{Riegler_transm},
a novel scheme for cross-talk compensation is introduced. In sections \ref{section3} and \ref{section4}
we discuss the generalization
of this concept to lossy N-strip systems. For that, in section \ref{section3} we identify the
dielectric losses by
measuring the RPC frequency response with a large bandwidth network analyzer,
giving a simple theoretical prescription on how to include them into the loss-less solutions. Following, in section
\ref{section4}, an extension of the aforementioned compensating scheme is given for N-strips, with particular
 focus on the literal solutions for 5-strip structures.
 A discussion on the results and the conclusions of the work are given in sections \ref{section5} and \ref{section6}.

\section{Cluster sizes in multi-strip RPCs}\label{section1}

Large cluster sizes either restrict the track identification capability of a time-of-flight detector or
proportionally increase its cost.
This is so because conceptual designs of this type of detectors, unlike tracking detectors, are usually based on
the average system occupancy as their main figure of merit. The latter is nothing else but the average
 probability that a detection cell/strip is hit per event, thus the ratio of the average number of tracks per event to the number of
  active cells $\bar{n} = N_{tracks}/N_{cells}$. For a
Poisson-distributed track multiplicity, the fraction of cells per event over which more than one track will imping can be calculated simply as:
\beq
P_{>1}=1-P_0-P_1=1-(1+\bar{n})e^{-\bar{n}} \label{clus1}
\eeq
assuming that each track produces a measurable signal in only one cell. The probability of
finding a cell with more than one hit ($P_{>1}$)
amongst all the fired cells is usually referred to as the `double hit probability':
\beq
P_{double}=\frac{P_{>1}}{P_1+P_{>1}} = \frac{1-(1+\bar{n})e^{-\bar{n}}}{1-e^{-\bar{n}}}\simeq \bar{n} \label{clus2}
\eeq
where the assumption $\bar{n}\ll 1$ has been made to derive the right-hand-side identity.
Very often it is not possible to identify any of the
two tracks arriving at the same cell so $2P_{double}$ represents, approximately, the
contamination from wrongly identified tracks that unavoidably
goes into the physics analysis (or the track-matching inefficiency, in case
tracks hitting those cells are discarded). Once an acceptable value
for $P_{double}$ is fixed by the physics goals (for instance, 5\%), preliminary cost
estimates can be performed (in approximate linear proportionally with the number of required cells).

In the more realistic case where each track causes in average a certain number of strips to fire ($\bar{n}_{s}$),\footnote{This is,
incidentally, the definition of cluster size.} eq. \ref{clus2} must be replaced by:
\beq
P_{double} = \frac{1-(1+\frac{N_{tracks}}{N_{cells}}\bar{n}_{s})
e^{-\frac{N_{tracks}}{N_{cells}}\bar{n}_{s}}}{1-e^{-\frac{N_{tracks}}{N_{cells}}\bar{n}_{s}}} \label{Pdouble}
\eeq
The number of cells required for keeping $P_{double}$ close to a certain design value in an
environment with a given number of tracks increases, thus, in direct proportionality
with the cluster size, unless an increase of $P_{double}$ as a function of $\bar{n}_{s}$,
according to eq. \ref{Pdouble}, is accepted.
Besides the shear occupancy problem, a track crossing a timing RPC may affect more than just
$\bar{n}_{s}$ strips. The electromagnetic perturbation on the strips potential, even if below
threshold, can affect the time measured for a coincident track in an uncontrolled way. This
possibility has been studied so far, perhaps strikingly, only for single-strip structures \cite{Blanco},
\cite{HADES_1}
where the effect should be absent. Even in these almost ideal conditions, a slight degradation of
the time resolution could still be seen in \cite{Blanco} affecting the first neighbors' performance under multi-hit
conditions, despite the cross-talk levels were as small as 0.4\%.

A priori, the most obvious candidate for increasing the RPC cluster sizes is the spatial spread
of the avalanche charge. Although no experimental value for the transverse diffusion coefficient $D_{_T}$ exists
for the standard RPC gas mixture, an estimate can be made based on the one recently measured for
the longitudinal diffusion coefficient $D_{_L}$
in pure C$_2$H$_2$F$_4$ \cite{Urquijo}
by using the ansatz $D_{_T}\simeq D_{_L}$, that is a good approximation at high fields. This yields a typical
avalanche diffusion radius $r = \sqrt{D_{_T}g/v_d} \simeq 10 \mu$m for a gap $g=0.3$mm under $E=100$kV/cm
(and $r \simeq 30 \mu$m for $g=2$mm, $E=50$kV/cm).
Given the typical scale of the read-out strips ($\sim$cm), the avalanche diffusion can
be thus expected to have a very minor role, indeed, in the observed cluster sizes.

Experimentally first \cite{ammosov} and latter theoretically \cite{Riegler_transm},\cite{Riegler_Ew},
\cite{Riegler_ind1}, \cite{Riegler_ind2}, the two main electrostatic effects that can
dominate the cluster sizes and the signal shapes in multi-strip RPCs have been identified.
Following \cite{Riegler_ind2}, in the next
subsection we discuss, semi-quantitatively, cluster sizes originated from the induction process.

\subsection{Charge-sharing}\label{charge_sharing}

\begin{figure}[ht!!!]
 \centering
 \includegraphics*[width=\linewidth]{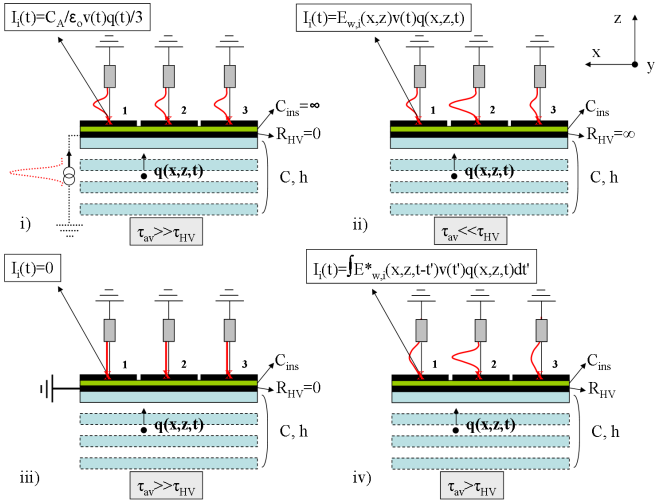}
 \caption{The four cases discussed in this section regarding signal induction: i) The HV coating behaves like
 an ideal conductor, yielding equal charge sharing among all strips, ii) the HV coating behaves like
 an ideal insulator, and the induction profile can be obtained from a standard static weighting field
 calculation  (units $[L^{-1}]$), iii) the read-out strips are fully shielded, iv) the presence
 of the HV coating has to be included for evaluation of the induction process yielding time-dependent
 weighting fields (units $[L^{-1} T^{-1}]$).}
 \label{cluster_size}
 \end{figure}

In the most general case nowadays, where the high voltage (HV) is applied through
a low-conductive coating of surface resistivity $R_s$ (for instance, \cite{ALICE}), four paradigmatic
situations can occur from the point of view of signal induction (Fig. 1):

\begin{enumerate}
\item The characteristic avalanche duration time $\tau_{av}\!\simeq\!g/v_d$ is much larger than the
response time of the HV coating $\tau_{_{HV}}$ under which the latter behaves in practice like a perfect
conductor ($\tau_{av}\!\simeq\!1.5$ns for $g=0.3$mm, $E=100$kV; $\tau_{av}\!\simeq\!18$ns for $g=2$mm, $E=50$kV,
as from \cite{Urquijo}). The induction can be seen as taking
place on the HV electrode (see left dotted current generator in Fig. \ref{cluster_size} up-left)
that is capacitively coupled to the read-out strips in a high-pass configuration.
When the coupling between HV and readout strips is ideal ($C_{ins}=\infty$) all the strips see
the same signal, proportional to the counter capacitance per unit area $C_{_A}$ divided by the number
of strips.
\item The characteristic  avalanche duration time $\tau_{av}$ is much smaller than $\tau_{_{HV}}$.
The avalanche induces currents
in the electrodes according to the `weighting fields' $E_{w}$ obtained upon application of the
Ramo theorem \cite{Ramo}. The weighting fields determined in that way
can still, in specific geometries, cause sizeable cluster sizes.
\item For completeness, an abstraction can be made on case i), by imposing the additional condition that the
coating represents an ideal ground (a situation not obviously realizable in practice). In this case,
the signal disappears at the coating, by definition, thus screening the readout strips completely.
\item At last we can consider the avalanche duration to be just larger than $\tau_{_{HV}}$
(intermediate situation). A time-dependent weighting field must be then calculated. A simple
prescription is given in \cite{Riegler_ind2} on how to do this, together with
practical examples for 1-gap chambers. As simple as it may be, no attempt has been made, so far, to
verify this model.
\end{enumerate}

\begin{figure}[ht!!!]
\centering
\includegraphics*[width=4.3cm]{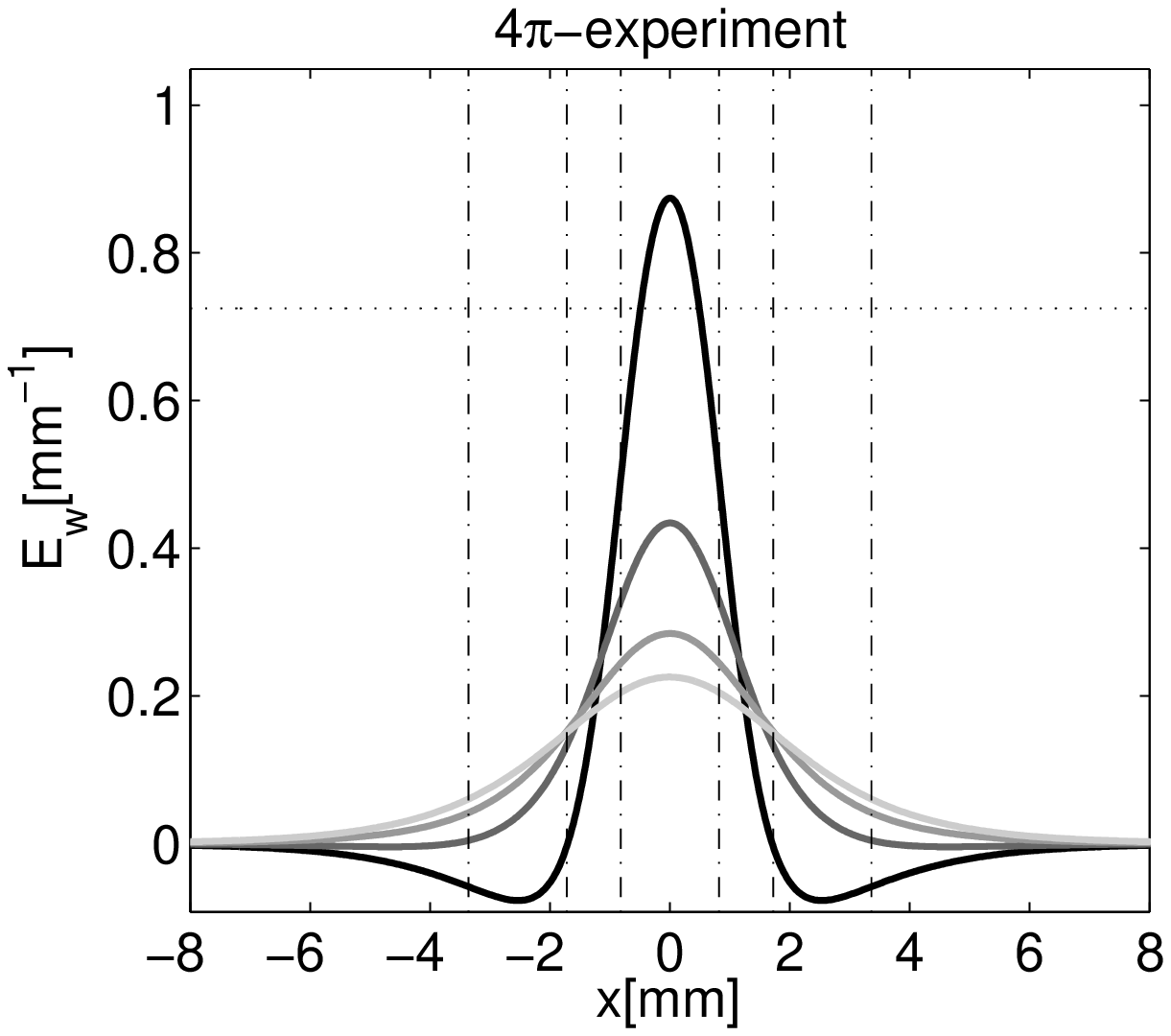}
\includegraphics*[width=4.3cm]{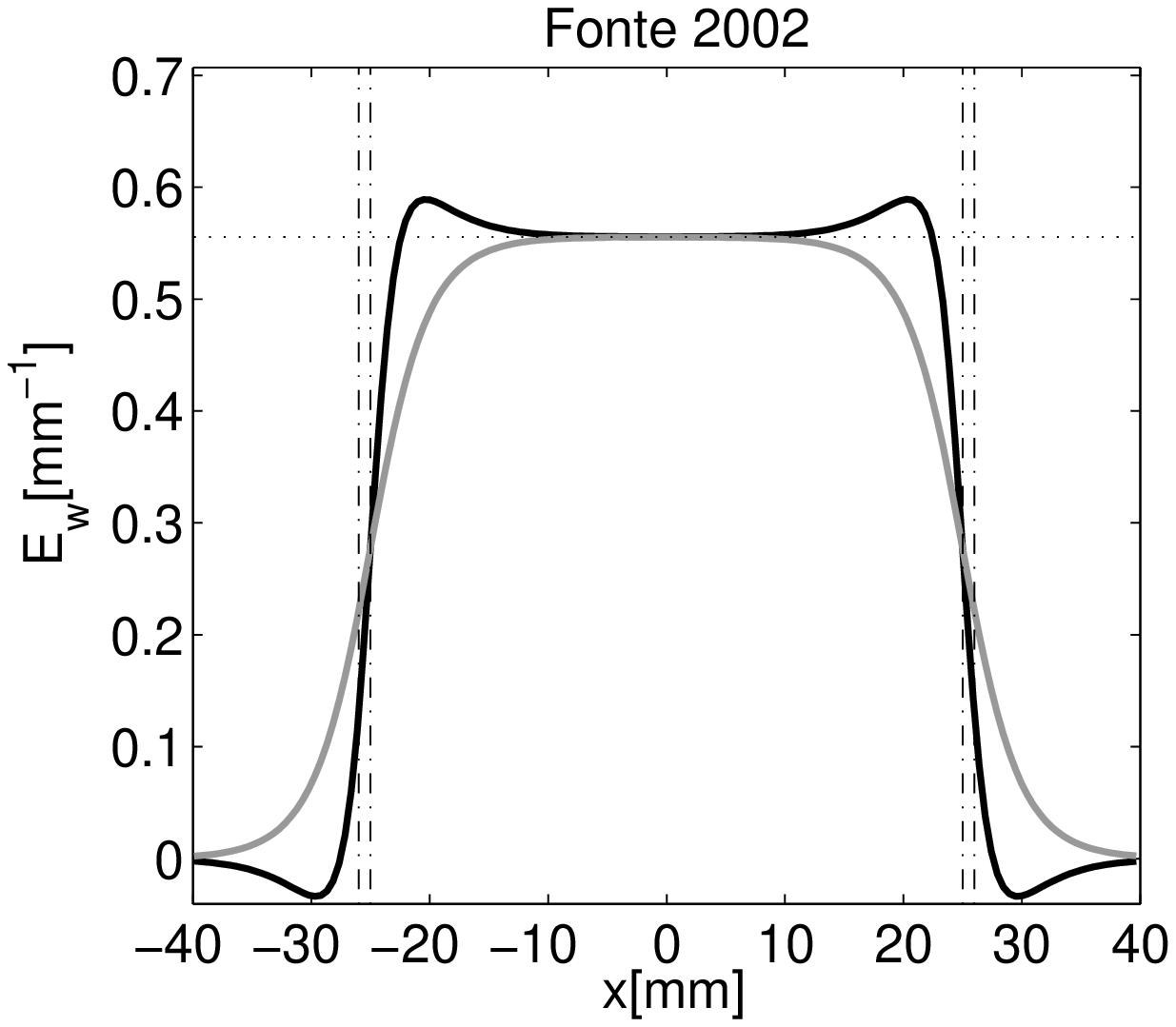}
\caption{Weighting field profiles across the strips ($x$ coordinate),
as obtained for a strip centered at 0,
evaluated at the center of each
gap, for the $4\pi$-experiment (left) and
Fonte-2002 prototype (right). Lines shown in gray gradient from black (gap closest to the read-out strip)
to light gray (gap furthest from the read-out strip). The dashed line shows $E_w=C_{_A}/\epsilon_0$.}
\label{Ew_profile}
\end{figure}

It is important to realize an implicit assumption made in \cite{Riegler_ind2}, and thus in
cases i)-iv): the region affected by the induction process $r_{ind}$ must be electrically short
 $r_{ind}\leq5$cm (from eq. \ref{Dstar2}). Otherwise, field propagation must be
obviously taken into account. This fact highlights the difficulty of addressing 
cases i) and iii) in a practical situation.\footnote{For instance, the theorems obtained
in \cite{Riegler_ind2} predict 
an equal signal sharing for infinite strips placed over an infinite floating conductor. However,
 any signal registered on the strips placed at infinite can only arrive after an infinite time, since 
the induction can not take place faster than the speed of light.}
For the geometries discussed here, however,
and for most practical cases, it is reasonable to assume that $r_{ind}\leq5$cm (see
for instance the weighting field profiles obtained in two extreme scenarios in 
Fig. \ref{Ew_profile}).

Under the, usually desired, condition $\tau_{av}\!\ll\!\tau_{_{HV}}$, the cluster size originated
during induction is minimal and can be obtained from a
static weighting field profile determined with the Ramo theorem alone (case ii)). This situation
is often given in practice. In order to better understand the implications of this condition, 
we take the expresion for $\tau_{_{HV}}$ obtained in \cite{Riegler_ind2}:
\beq
\tau_{_{HV}} \simeq R_s \epsilon_0 h \label{t_HV}
\eeq 
where $h$ can be interpreted as the anode cathode-distance and $R_s$ is the surface
resistivity of the material in [$\Omega/\square$].\footnote{This
expression is exact only for 1-gap chambers without resistive plates, for which the problem could be analytically
solved in \cite{Riegler_ind2}. Since i) the time-dependent component of the weighting field is
exponentially suppressed with $\tau_{_{HV}}$ \cite{Riegler_ind2}, ii) the 
dielectric constant of the resistive plates
 is at most a factor $\times 10$ higher than the one of the gas gap, and iii)
the present discussion is based on orders of magnitude,
the approximate expression for $\tau_{_{HV}}$ in eq. \ref{t_HV} is kept.}
We take, for illustration, a typical value of $h=4$mm, that corresponds to a 2mm-bakelite/2mm-gap trigger RPC
and an 0.5mm-glass/0.3mm-gap 6-gap timing RPC. Thus, the condition $\tau_{av}\!\ll\!\tau_{_{HV}}$
requires values for the resistivity of the
coating $R_s>500$ k$\Omega/\square$ and $R_s>50$ k$\Omega/\square$ for trigger and timing RPCs, respectively. 
Coatings at the scale of 10 M$\Omega/\square$ (for instance \cite{CBMingo}) seem to be thus very fitting.

For
illustration, weighting fields for two distinct RPCs are calculated as in \cite{Diego}.\footnote{In these two detectors
the signals are directly read-out from the electrodes on HV-potential. The HV plane
does not cause therefore signal screening, and the situation is analogous to
case ii) in Fig. \ref{cluster_size}.}
These geometries are characterized by two extreme values of the ratio of strip width to anode-cathode distance:
$\tn{w}/h\simeq 0.3 \ll 1$ \cite{FOPI} and $\tn{w}/h\simeq 5 \gg 1$ \cite{Fonte},
generating thus very different profiles across the strip, as shown in Fig. \ref{Ew_profile}.

In order to determine the region over which a strip can effectively `see' an avalanche
signal (i.e., the signal being above threshold) a full simulation is
needed \cite{Diego}. This approach is, however, outside the scope of the present work.
An approximate idea of the size of the induction phenomena can be 
obtained by calculating the region $\Delta{X}$ over which the average weighting 
field drops to a large fraction of its maximum value ($10\%$, for instance).
The area of influence of a strip beyond its geometrical limits can be estimated 
after substracting its width: $r_{ind}=\Delta{X}-\tn{w}$. 
This yields $r_{ind} \simeq 1.4{\tn{pitch}}$ for \cite{FOPI}
and $r_{ind} \simeq 0.1{\tn{pitch}}$
for \cite{Fonte}.
Reciprocally, `how far a strip can see' is directly related to `from how
far an avalanche can be seen', thus to the cluster size.
Intuitively, considerations on the resistance of the HV layer apart, the condition $\tn{w}/h\gg 1$ is thus
expected to minimize the cluster size originated during induction. A detector not fulfilling
$\tn{w}/h\gg 1$ may additionally
increase its cluster size by mere geometrical considerations. E.g., assuming an
angle of incidence in the plane transverse to the strips of $\theta=30$deg with respect to perpendicular incidence,
the track projection over the strips plane would be $r_{ind,\theta}\simeq 2h\tan{\theta}$
yielding $r_{ind,\theta}=2.3{\tn{pitch}}$ for \cite{FOPI} and $r_{ind,\theta}=0.22{\tn{pitch}}$ for \cite{Fonte}.

Although negative weighting fields can yield induced signals of opposite polarity, specially
for avalanches in the region between the strips and close to them (Fig. \ref{Ew_profile}),
the net signals originated during induction have generally the same polarity, thus constituting effectively
an area over which the avalanche-induced charge is `spread'.
We will refer to this phenomena as `charge sharing' (or `avalanche foot-print'),
 to make clear the different underlying principle with respect to 
the main phenomena of interest in this work, that is
introduced in the next section.

\subsection{Cross-talk}

\begin{figure}[ht!!!]
\centering
\includegraphics*[width=7cm]{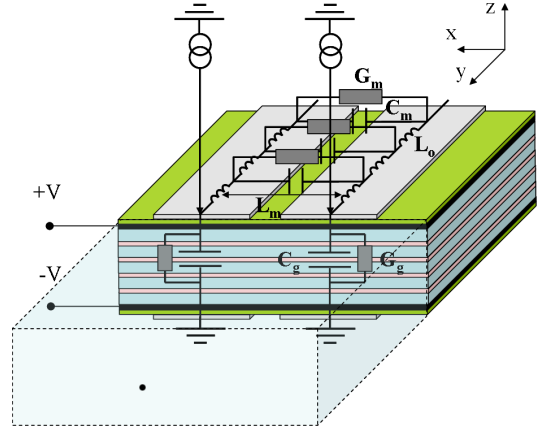}
\caption{Illustration of the transmission problem in 2-strip RPCs. The
currents may be obtained with the formalism of section \ref{charge_sharing}, for instance.}
\label{transm}
\end{figure}

According to \cite{Riegler_ind2} the finally measured currents and voltages can be determined once the
currents induced at the electrodes have been calculated (as in Fig. \ref{cluster_size}) but only after
introducing all the resistive, capacitive and inductive elements present in the system, including
the RPC itself. The associated circuit problem must be then solved using as input 
the calculated currents modeled as ideal generators. In case of being electrically-long, however, 
an RPC can not be characterized by conventional circuit theory 
and a distributed circuit theory is needed, where electrostatic elements
per unit length are used as input parameters (see Fig. \ref{transm} for a simple 2-strip situation).
The argument sketched
above suggests a separation between the transverse and longitudinal signal dynamics
and will be followed here for the sake of simplicity, as in \cite{Diego}. Going beyond this assumption
requires of a 3-dimensional modeling of the structure.

We discuss first the simpler case of signal transmission in loss-less structures (only capacitive and
inductive elements are present). As shown in the next section, the main transmission patterns are indeed
emerging from the structure of the solutions to this problem. 
A dedicated discussion on dielectric losses (due to the
shunting conductances $G_g, G_m$ in Fig. \ref{transm}) is postponed to section \ref{section3}. The
presence of skin effect would yield additional resistive elements in Fig. \ref{transm} but it is shown
(also in section \ref{section3}) to be a minor effect and has been neglected for the sake
of providing a simpler image of the process.

The analytical solutions of the N-strip transmission line equations
(often named `the telegrapher's equations') for the loss-less case
can be found, for instance, in the excellent overview book of C. R. Paul \cite{Clayton}.
They were introduced in the Resistive Plate Chamber field by W. Riegler in 2002 \cite{Riegler_transm}
in the context of impedance-matched systems. A
straight-forward generalization of the formalism in \cite{Riegler_transm} in order to explicitly
take into account the reflections was given in \cite{Diego}
for the 2-strip case. For the N-strip case the complete solution can be compactly written
as:\footnote{In the following, a variable with hat $\hat{}$ denotes an $N \times N$ matrix.}

\beq
\vec{I}_{T}\!(t)\!\! = \!\!\frac{\hat{Z}_{in}}{R}\!\frac{\hat{T}}{2} \!\! \sum_{j=0}^{\infty}\! (1\!-\!\hat{T})^j\!
\hat{M}\!\!\left(\!\!\begin{array}{c} \hat{M}^{-1}_{1n}\!I(t\!-\!\frac{(\!-1\!)^j y_0 \! + \! 2\lceil{j/2}\rceil\!{D}}{v_1})
\\\ldots\\ \hat{M}^{-1}_{Nn}\!I(t\!-\!\frac{(\!-1\!)^j y_0 \! + \! 2\lceil{j/2}\!\rceil{D}}{v_N})  \end{array}\!\!\right) \label{full_blast}
\eeq
where $\vec{I}_{T}(t)$ is the $N$-dimensional array of currents measured through the corresponding resistances $R$
placed at $y=0$ when the N-strip structure is excited along line $n$ by a current
$I(t)$ originated at position $y=y_0$.\footnote{We assume that all ports
are read-out with the same electronic system, having an input resistance R. In the most general case where several read-out
systems are present $R$ should be replaced by a vector.} The sum extends over all $j$ reflections and $\lceil j/2 \rceil$ denotes the next higher integer of $j/2$. $\hat{T}$ is the (in-out) transmission coefficient of the line:
\beq
\hat{T}=2 \hat{Z}_c (\hat{Z}_{in}+\hat{Z}_c)^{-1}
\eeq
$\hat{Z}_c$ is the characteristic impedance matrix of the transmission line
and $\hat{Z}_{in}$ is the impedance matrix with which it is loaded, that must be calculated
according to \cite{Riegler_transm}. In the most typical case where all ports are terminated with the
amplifier input resistance $R$:
\beq
\hat{Z}_{in} = \hat{1}R
\eeq
and the pre-factor in eq. \ref{full_blast} becomes 1. $\hat{M}$ is the matrix of eigenvectors with elements $\hat{M}_{ij}$ and $\vec{v}=\{v_1,\ldots,v_{_N}\}$ the array of
the inverse squares of the eigenvalues of the following diagonalization problem:
\beq
\hat{M}^{-1} (\hat{C}\hat{L}) \hat{M} = (\hat{1}\vec{v})^{-2} \label{dia_pro}
\eeq
with the impedance matrix being defined as:
\beq
\hat{Z}_c=\hat{L}\hat{M}(\hat{1}\vec{v})\hat{M}^{-1} \label{Z_exact}
\eeq
The notation $\hat{M}^{-1}_{ij}$ indicates the element with column/row indexes $i/j$ of the
inverse matrix $\hat{M}^{-1}$. $\hat{L}$ and
$\hat{C}$ are the `per unit-length' inductance and capacitance matrices of the structure
and $\hat{1}$ is the unit-matrix. Under the (most usual)
situation where all materials have a relative magnetic permeability $\mu_r\simeq 1$,
the following relation applies \cite{Demo_L_vs_C}:
\beq
\hat{L}=\frac{1}{c^2}\hat{C}_0^{-1} \label{LvsC}
\eeq
Here $\hat{C}_0$ is the capacitance matrix of the transmission line with all dielectrics replaced
by empty space, an equivalence that will be used throughout this document.
A description of the structure of the matrices $\hat{C}$ can be found in section \ref{section4}
or in \cite{Clayton}. It must be noted that for parallel-plate transmission lines, even in
case of a homogeneous surrounding medium, analytical calculations of the elements of the
capacitance matrices exist in just few cases.
For a parallel-plate structure that is also inhomogeneous, a numerical solution is enforced.

There are some relevant properties that emerge only after including the reflections
explicitly (as in eq. \ref{full_blast}), and
the most evident ones have been discussed in \cite{Diego}, among them the `delayed cross-talk' (the
fact that cross-talk stemming from a reflection at the opposite strip-end can largely exceed the direct
cross-talk) and the charge conservation (meaning that the charge induced in the main strip is collected
after summing up all the reflections and, under the same conditions, cross-talk does not transport net charge between
the strips either), a statement that is proved in the appendix of this work. This fundamental property of cross-talk in
loss-less lines (the absence of net charge when integrating over a large time window) has been used in \cite{Fonte_long}
in order to experimentally demonstrate its importance in multi-strip RPCs.

Eq. \ref{full_blast}, in spite of being analytical, is not very useful in
its present form. The long and tedious algebraic computation can be
carried out in a symbolic way, with Mathematica \cite{Mathematica} for instance,
but it is very difficult to grasp the meaning of the multiple terms arising and how an optimization
can be realized in practice. Solutions are often just too general, while the particular application
my fall easily under a set of reasonable simplifying conditions, making the solutions
of eq. \ref{full_blast} more enlightening as well as the fundamental variables ruling the phenomena.
Following \cite{Clayton}, we will call this type of solutions `literal' solutions. Literal solutions for an
inhomogeneous un-matched 2-strip line can be found, for instance, under the
low-coupling approximation \cite{clayton_2strip}, \cite{Diego} and their properties
are discussed in the next section.

\section{A systematic study of cross-talk and signal integrity in 2-strip counters}\label{section2}

\subsection{Introduction to the problem}\label{section21}

It has been recently reported \cite{Diego} that modal dispersion could be responsible
for the extreme cross-talk patterns observed in early implementations of timing
Resistive Plate Chambers with multi-strip readout \cite{Fonte_long}. Modal dispersion emerges from
the structure of the solutions of N-conductor lossless transmission lines (eq. \ref{full_blast})
when transmission is performed through inhomogeneous dielectric structures.
As a matter of fact, an RPC is intrinsically an inhomogeneous transmission line.
This very relevant feature can not be altered in view of the
simultaneous need of amplifying gas ($\epsilon_r\!\!=\!\!1$) and HV insulator
(either float glass or Bakelite, $\epsilon_r\!\!=\!\!5$-$10$).

\begin{figure}[ht!!!]
 \centering
 \includegraphics*[width=8.7 cm]{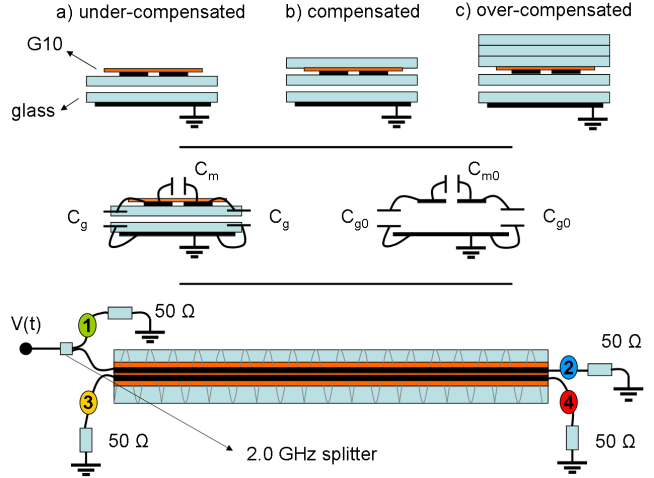}
 \caption{Up: Transverse section of the three typical cases studied in this work, from left to right: a system with
negative velocity dispersion
(under-compensated), zero (compensated) and positive (over-compensated). Middle: description
of how to calculate the electrostatic parameters needed for solving the loss-less transmission line problem.
Down: Up view of the scheme used for the measurements presented in this section. 
A fast pulser signal $V(t)$ (output resistance $50\Omega$) is connected to the left
of the structure 
via a 2GHz-splitter (port 1). All 4 ports are measured simultaneously with a 2.5GHz scope and terminated
with $50\Omega$. For the measurements performed with a 3.5GHz 4-port network analyzer, the connections
were made directly to the structure, without additional elements.}
 \label{scheme}
 \end{figure}

In the following we will assume that the reader is familiar with the literal solutions
to the exemplary 2-strip problem in the form presented in \cite{Diego}. They represent a particular
case of the general N-strip situation in eq. \ref{full_blast}, where the diagonalization
problem is particularly easy. Following \cite{Diego}, the subsequent discussion can be
stream-lined by recalling the `2-strip parameters':
\begin{enumerate}
\item Propagation velocity $\bar{v}$ (average velocity of the two system modes).
\item Velocity dispersion $\Delta v/\bar{v}$ (relative velocity difference of the
two system modes).
\item Characteristic impedance $Z_c$ (diagonal element of the characteristic impedance matrix).
\item Coupling coefficient $Z_m/Z_c$ (ratio of the non-diagonal element of the impedance matrix
to the diagonal one).
\end{enumerate}
A deeper insight can be obtained by recalling their approximate expressions under the
condition $Z_m/Z_c\!<\!1$ (low coupling):
\bear
 &&\bar{v}\simeq\sqrt{\frac{C_{g0}+C_{m0}}{C_g+C_m}}c, ~~~ \frac{\Delta v}{\bar{v}}\simeq \frac{C_{m}}{C_{g}+C_{m}}-\frac{C_{m0}}{C_{g0}+C_{m0}} \label{par1} \\
 &&Z_c\simeq \frac{1}{\sqrt{(C_{g0}+C_{m0})(C_{g}+C_{m})}}\frac{1}{c} \label{par2} \\
 &&\frac{Z_m}{Z_c}\simeq \frac{1}{2}\left(\frac{C_{m}}{C_{g}+C_{m}}+\frac{C_{m0}}{C_{g0}+C_{m0}}\right) \label{par3}
\eear
 They depend on the capacitance with respect to ground $C_{g}$
and the mutual capacitance $C_m$ per unit length both in the real structure and in the
empty space ($C_{g0}$, $C_{m0}$) and can be obtained by solving the corresponding 2-dimensional
 electrostatic problems for a cross-section of the device (Fig. \ref{scheme}-middle).
In the limit $C_m,C_{m0}\rightarrow0$ the well-known 1-strip (1 conductor + 1 reference)
expressions for $\bar{v}$ and $Z_c$ are recovered
by recalling that the induction coefficient per unit length is then $L_0=1/(c^2C_{g0})$.
Signal transmission and cross-talk in a 2-strip line (2 conductors + 1 reference) can
be expressed conveniently as a function of the parameters above, as shown in \cite{Diego}.
Other descriptions of the loss-less 2-strip situation are however possible: as an example, classical
circuit models based on odd and even impedances
and velocities exist since long time \cite{zysman}, while more tractable literal solutions
for a general un-matched case (similar to \cite{Diego}) can be found
more recently (\cite{clayton_2strip}, for instance). An approximate, although insightful,
derivation can be found in \cite{Feller} and references therein.

In an inhomogeneous structure the velocity/modal dispersion 
given by eq. \ref{par1}-right can dominate the cross-talk and transmission
patterns well beyond the shear strength of the electrostatic
coupling $Z_m/Z_c$ (eq. \ref{par3}). Indeed, its importance depends critically on the
propagation distance and the signal rise-time.
As it can be deduced from eq. \ref{par1}, the velocity dispersion is zero
for a homogeneous material with arbitrary dielectric constant $\epsilon=\epsilon_0\epsilon_r$,
 since $C_{g}=\epsilon_r C_{g0},~ C_{m}=\epsilon_r C_{m0}$.
Not being this statement generally true for an inhomogeneous structure, the value of
$\Delta v/\bar{v}$ may be, however, `adjusted'. A
simple implementation of this idea is shown in Fig. \ref{scheme}-up: values in empty
space are not changing due to the presence
of an additional dielectric above the readout electrodes, the coupling to ground is virtually unaffected,
and only $C_m$ varies. The labels refer to 3 paradigmatic cases: `under-compensated' ($\Delta v/\bar{v}<0$),
`compensated' ($\Delta v/\bar{v}=0$) and `over-compensated' ($\Delta v/\bar{v}>0$). A system 
is thus said to be compensated
when the coupling coefficient $Z_m/Z_c$ is the same in the filled and in the empty structure, $Z_m/Z_c=Z_m/Z_c|_0$,
as can be deduced from eqs. \ref{par1} and \ref{par3}. The velocity dispersion of
the two system modes is then zero and, equivalently, the capacitive and inductive coupling are
 balanced $C_m/(C_g+C_m)=L_m/L_0$, through relation \ref{LvsC}. This symmetry was realized long ago,
but it is usually regarded as a feature proper only of homogeneous lines \cite{Feller}.
It is therefore very appealing to explore the possibility of constructing
an RPC fulfilling the condition $\Delta v/\bar{v}=0$. Such a compensated
system theoretically exhibits minimal signal shaping and cross-talk, being its properties independent from
the propagation distance, as long as losses can be neglected.

Several caveats to the above interpretation are worth being noted at this point. First, transmission
through inhomogeneous structures precludes a pure TEM-mode propagation (where the electric and magnetic fields
are transverse to the direction of movement) and thus formally invalidates the
telegrapher's equations and the solutions given in eq. \ref{full_blast}. Moreover, the loss-less assumption
that, if violated, also invalidates eq. \ref{full_blast}, was not proved for neither typical glass nor
Bakelite -based RPCs.
It is a common practice to include these aforementioned facts under the name `quasi-TEM' approach and
to use the telegrapher's equations
anyway, however a sound experimental measurement is then required. Eq. \ref{full_blast} is also
invalidated in the presence of frequency or direction-dependent electrical properties,
that is sometimes the case in dielectric materials. Last but not least,
sample-to-sample variations of the electrical properties, or simply the required mechanical
accuracy or its tolerance (that defines the line uniformity) may pose unrealistic requirements
for a practical realization.

We have therefore designed a high precision experiment aimed at proving both the dominant role
of modal dispersion in long timing RPCs and the possibility of implementing a simple compensation technique
(hereafter `electrostatic compensation').

\subsection{Electrostatic compensation in 2-strip structures}\label{section22}

\subsubsection{Description of the experiment}

Several 2meter-long electrodes were specially manufactured, each
consisting of 2 parallel $0.05$mm-thick, $25$mm-wide copper strips
on an $0.25$mm-thick epoxy glass laminate as substrate (G10). The strip width and inter-strip separation
were accurately defined within $\pm0.1$mm, as verified by microscope.\footnote{I.e, the standard
 deviation of the inter-strip distance as measured along the strips length is of the order of 100 $\mu$m.}
Electrodes not fulfilling this condition were rejected for the experiment.\footnote{The acceptance
yield was 4/20.}
The basic test structure was that of a micro-strip configuration with strips placed above a
1 or 2 -gap structure laying on a quasi-infinite ground plane (Fig. \ref{scheme}).\footnote{The width
of the ground plane and glass plates was 60 cm and 50 cm respectively. According to MAXWELL-2D
simulations (see later),
 they can be effectively considered to be infinite for the capacitance calculations.} Various
inter-strip separations were essayed, but
only 2.1mm and 3.1mm were systematically characterized. The definition of the gas gaps was performed
through nylon monofilaments of $0.3\pm0.01$mm diameter interleaved
on $1\pm0.05$mm float glass plates,\footnote{Schott.} arranged in the direction across the strips with a 5 cm pitch.
We present first the measurements performed in the time domain and a separate discussion
is devoted to high-precision frequency-domain measurements in order to assess losses (section \ref{section3}).
A pulser with rise-time $t_{rise}=280$ps,\footnote{Defined as the time elapsed from a fraction 0.1 to 0.9 of the signal maximum.}
FWHM $\Delta{t}=700$ps,\footnote{Full width at half maximum.} a repetition rate of 50 Hz
and output impedance $R=50 \Omega$
was injected into one of the ports (1) and
signals recorded in the opposite port (2), in the near-end (3) and in
the far-end cross-talk ports (4), (see Fig. \ref{scheme}-down).
The signal was split before injection with a 6dB-2GHz signal divider and
sent both to the electrode structure and to a 2.5 GHz Tektronics scope at 10 Gsamples/s.
Ten cm long $50\Omega$ BNC cables terminated on $50\Omega$ 
were attached both to the far-end of the structure
and to the near-end cross-talk port (3). 
The electrical connection between
those cables and the electrodes
was performed over 1 cm length through flexible copper strips soldered with tin.

The uniformity of the line was ensured by applying weight on 10 cm-thick blocks of extruded
polystyrene foam via stainless steel bricks. The foam was attached directly to the strips.
A high uniformity proved to be extremely important,
since the very thin electrodes tended to bend-up easily over 1mm or more and break the line
impedance, giving immediately very large cross-talk and dispersion patterns. We will ascribe
the electrical behavior of this auxiliary ensemble to that of air, with $\epsilon_r=1$. We have evaluated
both in experiment and simulation the effect of the foam thickness and the steel, and concluded that it can be
effectively considered to behave electrically like air.

Measurements were stored when the waveforms at all ports were
independent from additional pressure applied onto the strips and all connections had been checked.
Compensation was simply achieved by placing additional glass plates above the electrodes.

\subsubsection{Measurements}

As indicated in Fig. \ref{scheme}, all the measurements performed in the following were done
on a resistance $R=50\Omega$. For better representation, 
the fraction of transmitted signal $F_{tr}(t)$ is defined
as the ratio of the measured transmitted 
current $I_{tr}(t)$ to the maximum of the injected one $I(t)$. On the other hand, the fraction of
cross-talk $F_{ct}(t)$ is defined as the ratio of the measured cross-talk current $I_{ct}(t)$, 
to the maximum of the transmitted one:
\bear
&&F_{tr}(t)=I_{tr}(t)/\tn{max}[I(t)] \label{tr1}\\
&&F_{ct}(t)=I_{ct}(t)/\tn{max}[I_{tr}(t)]
\eear
where `max[]' denotes the maximum of the bracketed
function. We still need to decide how to define the injected signal $I(t)$
from the measured voltage in the scope. It turns out
that the following definition for the normalization of $I_{tr}(t)$ is very convenient:
\beq
I(t)=2 \frac{V(t)}{R} \label{ref_I}
\eeq
Here $V(t)$ is the voltage pulse measured after the splitter. 
As shown in appendix, under equivalence \ref{ref_I} the transmitted
signals correspond to the ones in the physical situation where a current $I(t)$ was induced at one
of the detector ends, $y_o=D$. Therefore, the solutions of eq. \ref{full_blast}:
\beq
\vec{I}_{_T}\!(t)\! =\! \frac{\hat{T}}{2}\! \sum_{j=0}^{\infty}(1\!-\!\hat{T})^j\!
\hat{M}\!\!\!\left(\!\!\begin{array}{c} M^{-1}_{11}I(t\!-\!\frac{(\!-1\!)^j\! D \! + \! 2\lceil{j/2}\rceil\!{D}}{v_1})
\\ M^{-1}_{21}I(t\!-\!\frac{(\!-1\!)^j\!D \! + \! 2\lceil{j/2}\rceil\!{D}}{v_N})  \end{array}\!\!\right) 
\label{full_blast_2}
\eeq
with $\vec{I}_{_T}(t)=\{I_{tr}(t),I_{ct}(t)\}$ can be directly compared with pulser data.
For that, the recipe to be followed is that $F_{tr}(t)$ from eq. \ref{tr1} is normalized according 
to eq. \ref{ref_I}. 

 \begin{figure}[ht!!!]
 \centering
 \includegraphics*[width=9.5 cm]{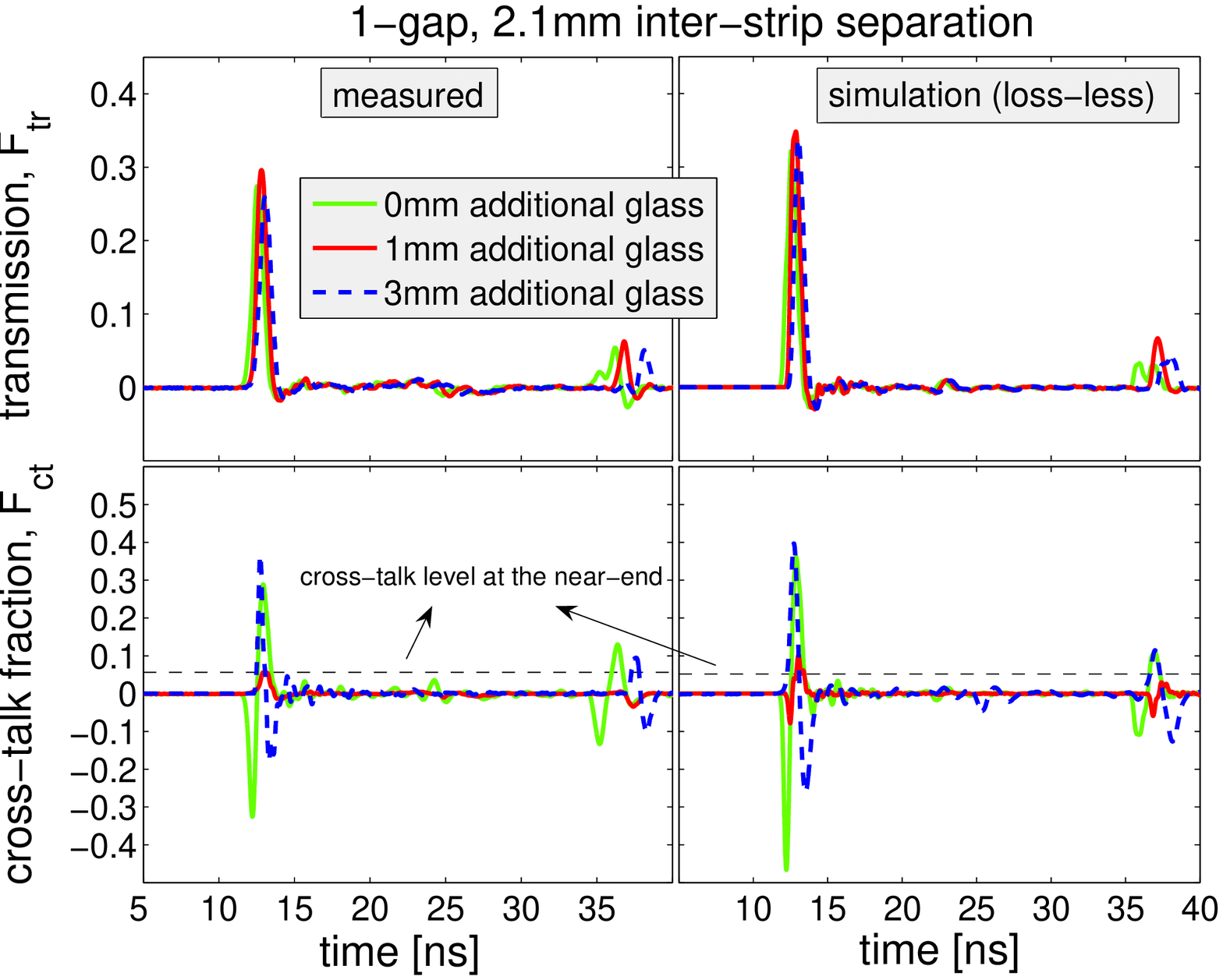}

 \includegraphics*[width=9.5 cm]{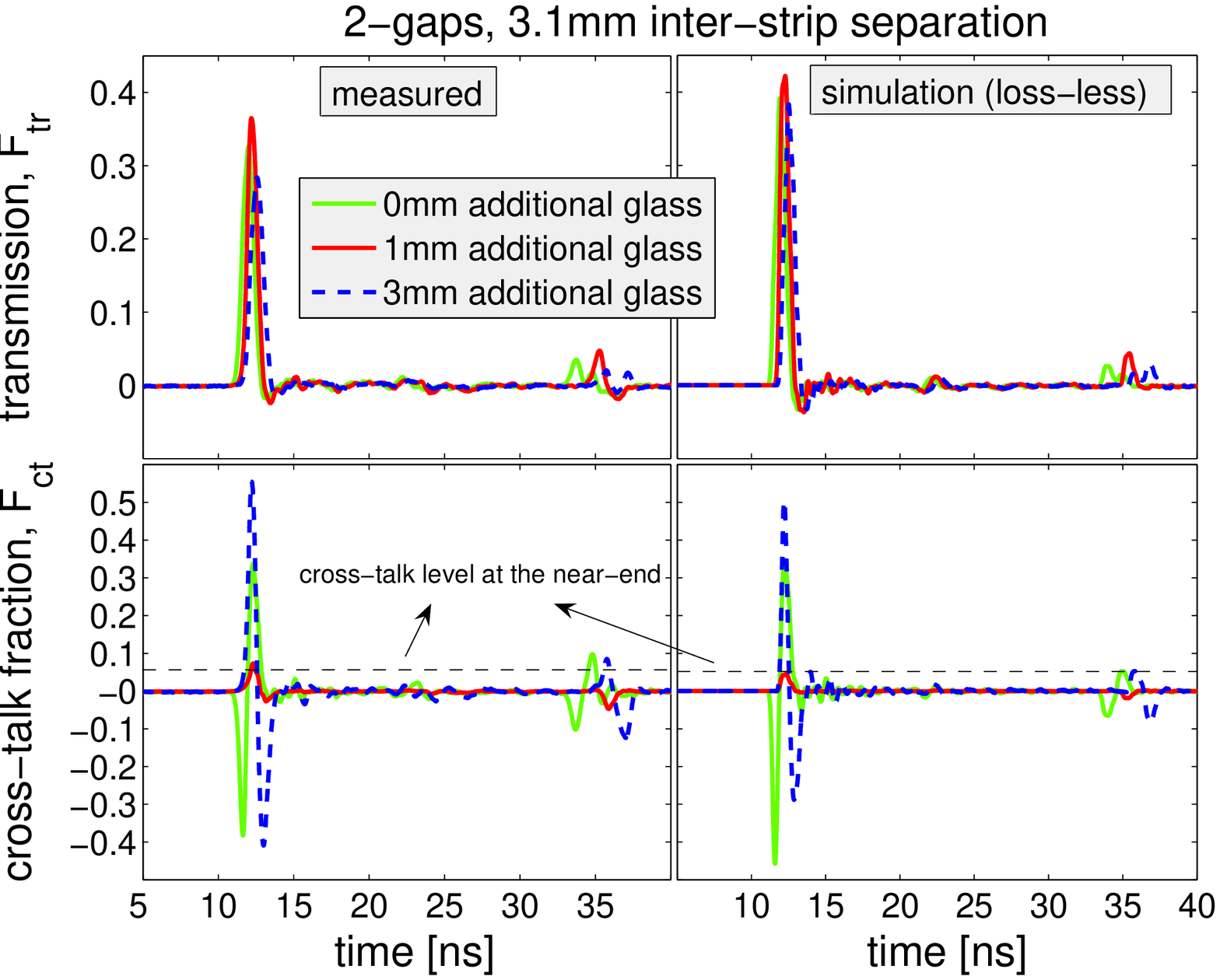}
\caption{Up-set: fraction of transmitted signal and cross-talk fraction for a 2-strip RPC with
 2.1mm inter-strip separation
  and 1 gas gap, for an under-compensated case (green/dark line), compensated (red/dark line) and
 overcompensated (dashed blue line). Low-set: as in the upper set, but for 3.1mm inter-strip separation
and 2 gas gaps. Up to a factor 10 cross-talk suppression
can be achieved if the system is adequately compensated. The far-end and near-end cross-talk
become equal in that case (horizontal dashed line). Figures to the right show simulations
assuming the structure to be loss-less and taking the measured value $\epsilon_r=5.5$ for float glass.
Capacitance matrices calculated with MAXWELL-2D.}
 \label{waveforms}
 \end{figure}

We present in Fig. \ref{waveforms} the oscillograms for the two cases for
which electrostatic compensation, $\Delta v/\bar{v}=0$, was achieved with the aforementioned prosaic
procedure of placing additional glass plates above the readout strips.
As shown in Fig. \ref{waveforms} up-left, for 2.1mm inter-strip separation
and 1 gas gap electrostatic compensation could be roughly achieved for 1 additional
glass plate (red/dark line), and similarly for 3.1mm inter-strip
spacing in case of 2 gas gaps, down-left.
The un-compensated systems show a distinct bipolarity with an additional difference in sign
from the under-compensated (green/gray line) to the over-compensated (blue dashed line) one. On the
other hand, the compensated systems show a factor 10 smaller cross-talk signal, having 
the same shape than the original one. Approximately equal levels are achieved in this case both for
the far-end and near-end cross-talk (horizontal dashed line in Fig. \ref{waveforms}).
In any case, after just including the first reflection, the cross-talk has
approximately zero net charge in any of these configurations. It approaches
the expected zero value in the limit where all reflections are included (see appendix). 

The overall behavior
of the oscillograms is reasonably captured (Fig. \ref{waveforms}-right) by a loss-less
simulation based on eq. \ref{full_blast_2}. Capacitance matrices were obtained
via a Finite Element Method (FEM) calculation from the MAXWELL-2D package \cite{MAXWELL}.
A relative dielectric constant for float glass
of $\epsilon_r=5.5$ was used, acccording to a direct measurement presented in section \ref{meas_glass}. 
For G10, a typical value $\epsilon_r=4.4$ was chosen.
The time-offsets present in the oscillograms due to cables were substracted in order to
match the simulated waveforms. Besides this fact, the simulations are therefore
parameter-free.\footnote{A full
Finite Difference Time Domain (FDTD) solution of the telegrapher's equations
performed with the APLAC HF-simulator \cite{APLAC}, used in \cite{Diego}, was also
attempted. The calculation shows, however,
numerical instabilities due to the presence of a large ground electrode and is being currently
investigated.}

 \begin{figure}[ht!!!]
 \centering
 \includegraphics*[width=9.5 cm]{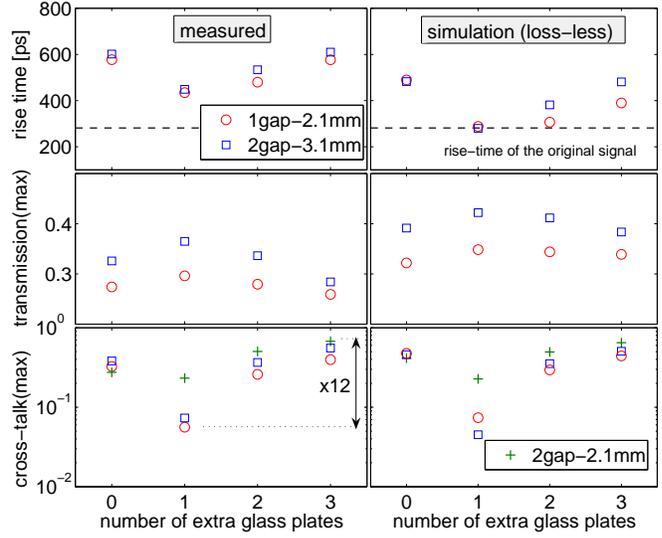}
 \caption{Left column: measured signal rise-time, maximum fraction of transmitted signal and maximum fraction
 of cross-talk for 2.1 and 3.1mm inter-strip separation for 2-strip 
structures with 1 ($\circ$) and 2 ($\square$) gaps,
respectively.
Right column: the same observables as obtained from a parameter-free simulation neglecting losses.
The compensated system ($\sim\!1$ additional
1mm-thick glass plate)
 clearly shows the most favorable properties. The 2gap-2.1mm case ($+$) is shown for further insight
(compensation could not be achieved since it requires a glass thickness in between 0 and 1mm, not available
during the measurements). Cross-talk can be suppressed up to a factor of 12 by simple design choices.}
 \label{exp_meas}
 \end{figure}

Despite the apparent close agreement in Fig. \ref{waveforms}, in order to perform
an adequate evaluation we have followed a more application-oriented approach: i.e.,
when aimed at a precise time determination at high efficiency in a multi-hit environment
the main figures of merit of a multi-strip system
are, probably: i) the deterioration of the
signal rise-time during transmission, ii) the
maximum transmitted signal ($\tn{max}[|F_{tr}(t)|]$) and
iii) the maximum crosstalk fraction ($\tn{max}[|F_{ct}(t)|]$). These quantities are compiled in Fig. \ref{exp_meas}
for measurements (left) and simulations (right) for various structures, as a function of the number
of additional glass plates.
The system clearly exhibits more favorable properties when it is compensated,
showing a higher transmission, lower shaping and minimal cross-talk.
The smaller measured transmission and larger signal rise-times
as compared to simulations could be traced back to losses in the line and are discussed
in detail in next section. It
may look like a small effect but the $\sim{200}$ps offset observed in data
with respect to simulations in Fig. \ref{exp_meas} up-left is about
a factor of two higher than the intrinsic signal rise-times expected for RPC signals
in the absence of Space-Charge, $t_{rise}=\ln 9 /S \simeq 110$ps. A precise
simulation of transmission patterns can be thus attempted only after including losses
and is given in the next section.

Crosses (+) in Fig. \ref{exp_meas}-down are aimed at
highlighting that compensation is not related solely to the usage of an additional
1mm-thick glass plate (for 2.1mm inter-strip separation and 2gaps compensation would require of
$\simeq 0.5$mm glass thickness, not available during the experiment). By looking at
the cross-talk patterns it is clear that the precision required for this system
to be compensated is well below 1mm. Precisely, 1mm difference, either in
the additional compensating glass or in the inter-strip distance, can easily imply
a cross-talk difference of up to a factor of $\times 10$, together with a worsening in
the signal rise-time by 200ps.

\subsubsection{The 2-strip parameters}

\begin{figure}[ht!!!]
 \centering
 \includegraphics*[width=9.5 cm]{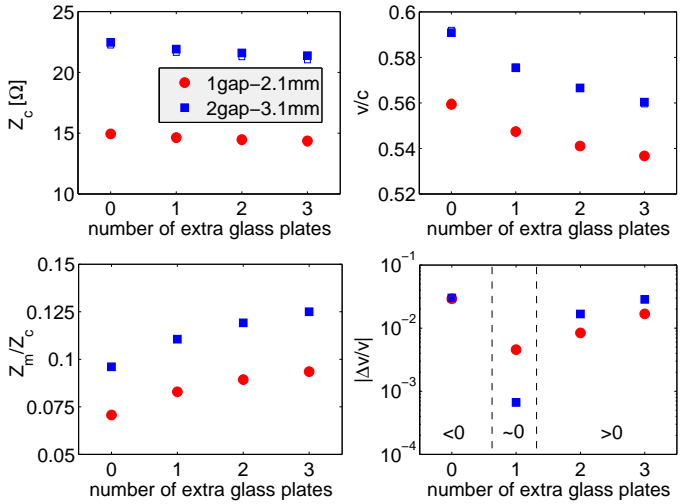}
 \caption{Simulated `2-strip parameters' for different number of gaps and two different
 inter-strip separations, from up-left to down-right: characteristic impedance,
 propagation velocity, coupling coefficient and velocity dispersion (absolute
 value). Values obtained from a FEM calculation performed with the MAXWELL-2D solver.
 Closed symbols show the exact value and open ones (almost undistinguishable) the value
 under the `low-coupling' approximation  (eqs. \ref{par1}-\ref{par3}).}
 \label{2strip}
 \end{figure}

The `2-strip parameters' contain all the information necessary for
characterizing a loss-less 2-strip
transmission line. They are shown in Fig. \ref{2strip}, as derived from the
capacitance matrices obtained from MAXWELL-2D.
Closed symbols indicate the exact values of the parameters while open symbols
(almost indistinguishable), show the values obtained under the `low-coupling'
approximation (eqs. \ref{par1}-\ref{par3}).

Note that the coupling coefficient increases steadily with the number of
additional glass plates, as intuitively expected, while the velocity dispersion shows
a shallow minimum for 1 glass plate, roughly corresponding to a compensated
situation.
The other parameters have a very smooth dependence with the number of glass
plates and clearly lack of importance for the phenomena here addressed.

\subsubsection{Solutions in the frequency domain for a loss-less 2-strip line}

\begin{figure}[ht!!!]
\centering
\includegraphics*[width=9.3 cm]{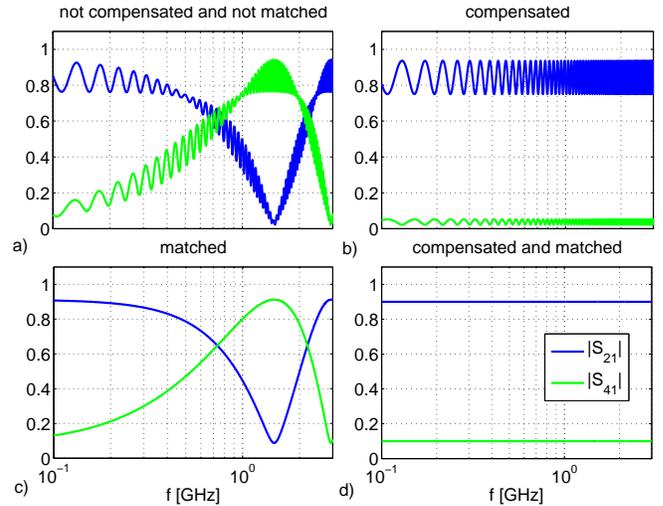}
\caption{Simulated moduli of the transmission $S_{21}$ and far-end cross-talk $S_{41}$ coefficients for
4 exemplary cases in a 2-gap/3-glass RPC with 3.1mm inter-strip separation.
From up-left to down-right: a) un-compensated and without impedance matching,
b) compensated but without matching, c) un-compensated but matched and d) compensated and matched.
No shaping is thus expected for the latter case, under the assumption of loss-less
line.}
\label{frec_res}
\end{figure}

Fig. \ref{frec_res} shows the simulated moduli of two
scattering matrix parameters $S_{21}$ (transmission)
and $S_{41}$ (far-end cross-talk) as obtained from the Fourier transform
of the time-domain solutions given in eq. \ref{full_blast} (for details see next section).
The base structure has an inter-strip separation of 3.1mm and 2-gap/3-glass
(as in Fig. \ref{waveforms}-down) for which the following cases were studied: a) un-compensated
(no additional glass) but un-matched, b) compensated (1.05m additional glass) but
un-matched, c) un-compensated but matched, d) compensated and matched. Note
that a 2-strip system requires of 3 resistors
in order to be perfectly matched, typically two in series with the ports and one
in parallel between ports on the same strip end ($R=24\Omega$ and $R_m=195\Omega$ here,
respectively). The oscillatory
pattern observed in Fig.\ref{frec_res}(up-right) is responsible for the reflections
observed in the counter while Fig.\ref{frec_res}(down-left) indicates the pure
effect of modal dispersion. The characteristic band-stop region at approximately
1GHz was first predicted and measured in micro-strips
by G. I. Zysman and A. K. Johnson as early as 1969 \cite{zysman}.\footnote{In printed circuit
design, micro-strips are usually inhomogeneous, unlike strip-lines.}
It must be noted that a homogeneous 2-strip loss-less system has a flat frequency response (no signal shaping)
if its impedance is matched while an inhomogeneous one requires, additionally,
to be electrostatically compensated.

\section{Deviation from the loss-less TEM propagation mode} \label{section3}

Losses have been neglected for the derivation of eq. \ref{full_blast} and Fig. \ref{frec_res} and have
been excluded from previous analysis \cite{Riegler_transm}. There
are two main sources of losses in a transmission line:

On the
one hand it stands the finite resistance of the readout strips at high frequency, since conduction
is then confined to a very thin layer (`skin effect'). Assuming that
conduction takes place along 75\% of a skin-depth (typical value, \cite{Clayton}),
the resistance per unit length of a strip of thickness $t\ll\tn{w}$ is, at high frequencies:
\beq
R\simeq \sqrt{\frac{\pi}{2} \mu \rho_{_{DC}}} \times \frac{\sqrt{f}}{\tn{w}} \label{skin}
\eeq
where $\rho_{_{DC}}$ is the DC conductivity of the material considered, and $\mu$
its magnetic permeability.

On the other hand, losses can be originated due to the shunting conductance between the strips themselves and/or
between strips and ground. Being usually very well insulated electrically (in RPCs either
glass or Bakelite can be effectively considered as ideal insulators for the sake of signal transmission), the shunting
conductance is expectedly governed by the dielectric losses on the insulator materials:
\beq
G_{glass} = 2\pi C_{glass} \tan{\delta}|_{glass} \times f \label{losses1}
\eeq
The loss-tangent $\tan{\delta}|_{glass}$ is the ratio of the imaginary to the real part of the
dielectric constant of the given medium and $C_{glass}$
is its capacitance with respect to ground (assumed to be glass here).
The losses due to either the polarization of air or
the standard gas mixture will be neglected in the following due to their very low
density of electric dipoles. However, even in this simple case, an analytical evaluation
of the losses becomes complicated in general for an inhomogeneous structure.
A practical approach to perform this calculation is given in \cite{Clayton}, by considering
that the (now) complex capacitance is given, to a good approximation, by the series capacitance
of the system (no edge effects). In such a case a simple parallel plate capacitor formula can be used
and, after some simple algebra, the conductance for the whole structure can be estimated as:
\bear
&&G \simeq 2\pi C_g F \tan{\delta}|_{glass} \times f \label{losses2} \\
&&G \simeq F G_{glass}, ~~ \tn{with} ~~~ F=\frac{C_g}{C_{glass}} \\
&&\frac{G}{G_{glass}} \simeq \frac{C_g}{C_{glass}}
\eear
where $C_g$ is the capacitance with respect
to ground of the whole structure. As intuitively expected, the shunting
conductance is thus reduced by interleaving gas. We will define the effective loss-tangent
of the structure as $\tan{\delta}^* = F\tan{\delta}|_{glass}$.

Losses are usually discussed in a frequency-domain representation, since they
are expected to have a
very characteristic dependence (\ref{skin}, \ref{losses1}). Losses can be experimentally
determined through a 1-strip transmission
measurement. For this it is useful to make use of the fact that 
the complex transmission coefficient from port $1$ to $2$ has the simple
analytical expression \cite{Clayton}:
\beq
S_{21}(f)=\frac{(2-T)T}{1-(1-T)^2e^{-2\gamma{D}}}e^{-\gamma{D}} \label{S21}
\eeq
For completeness, the reflection coefficient $S_{11}$ is given as:
\beq
S_{11}(f)=1-T\frac{1+(1-T)e^{-2\gamma{D}}}{1-(1-T)^2e^{-2\gamma{D}}} \label{S11}
\eeq
The transmission coefficient $T=2Z_c/(R+Z_c)$ is now a complex number, derived from the complex impedance $Z_c$:
\beq
Z_c=\sqrt{\frac{R+ j 2 \pi f L_0}{G+j 2 \pi f C_g}} \label{imped}
\eeq
and
\beq
\gamma=\sqrt{(R+ j 2 \pi f L_0)(G+j 2 \pi f C_g)}=\frac{1}{\Lambda} + j\beta \label{gamma}
\eeq

The reflection and transmission coefficients verify, in a loss-less system, the
condition $|S_{11}|^2 + |S_{21}|^2=1$.

Although the exact formulas \ref{S21}-\ref{gamma} will be used in the following for the sake of
precision, the `low-loss' approximation $\frac{G}{2\pi{f}C_g}, \frac{R}{2\pi{f}L_0}\ll 1$
is often used because at high frequencies it is
fulfilled for most practical purposes.\footnote{For a system dominated by
 dielectric losses this condition translates simply into $\tan{\delta}^*\ll 1$,
that is the case here as we will see, but also a rather typical situation for most
dielectric materials.} Under this approximation additional insight can be
obtained since:
\beq
\frac{1}{\Lambda} \simeq \frac{1}{\Lambda_{_G}} + \frac{1}{\Lambda_{_R}}
\eeq
being:
\bear
&&\Lambda_{_R} \simeq \frac{2 Z_c}{R} \label{L_R}\\
&&\Lambda_{_G} \simeq \frac{2}{G Z_c} \label{L_G}
\eear
and $\beta$ equals:
\beq
\beta \simeq \frac{2\pi{f}}{v} = 2\pi{f}\sqrt{L_0C_g}
\eeq
As it can be readily obtained from eqs. \ref{skin}, \ref{losses2}, \ref{L_R}, \ref{L_G}
the geometrical dependence of $\Lambda_{_R}$, $\Lambda_{_G}$ with w is canceled
in first order for wide-strip RPCs ($\tn{w}\gg h$ $\rightarrow$ $(Z_c, C) \sim 1/\tn{w}$).
So, in practice, the main variables ruling the losses are indeed
the frequency, the propagation distance and the loss-tangent. The
cutoff frequency for each of the two processes \ref{L_R}, \ref{L_G} can be obtained
approximately from the dominant $e^{-\frac{D}{\Lambda}}$ behavior in eq. \ref{S21}, yielding:
\bear
&&f_{c,R} \simeq \left(\frac{Z_c \tn{w} \ln{2}}{D} \right)^2 \frac{2}{\pi\mu\rho_{_{DC}}} \label{fc_R}\\
&&f_{c,G} \simeq \frac{v \ln{2}}{2\pi D \tan{\delta}^*}  \label{fc_G}
\eear

\subsection{Losses in 1-strip structures}\label{meas_glass}

\begin{figure}[ht!!!]
\centering
\includegraphics*[width=\linewidth]{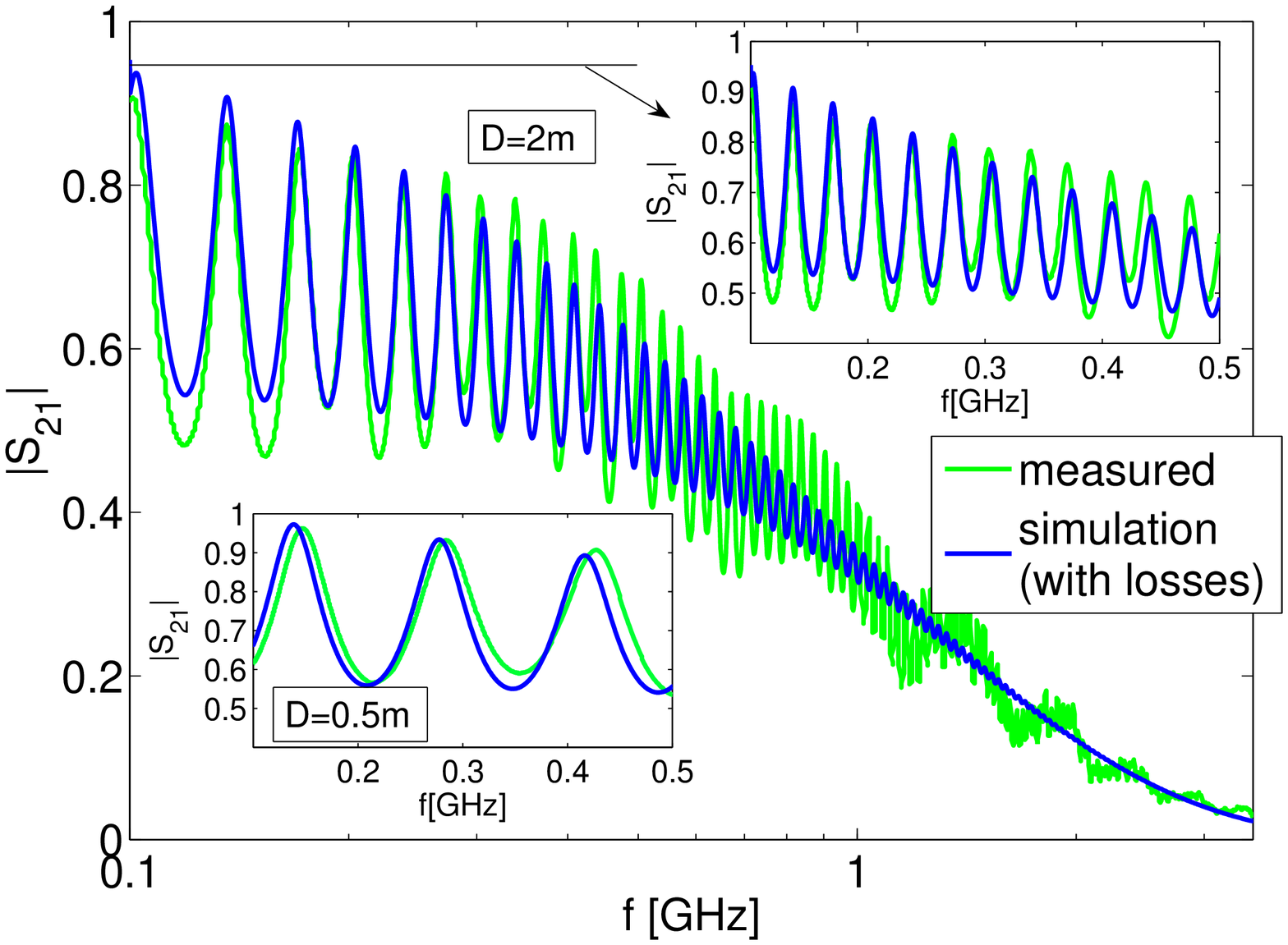}
\includegraphics*[width=\linewidth]{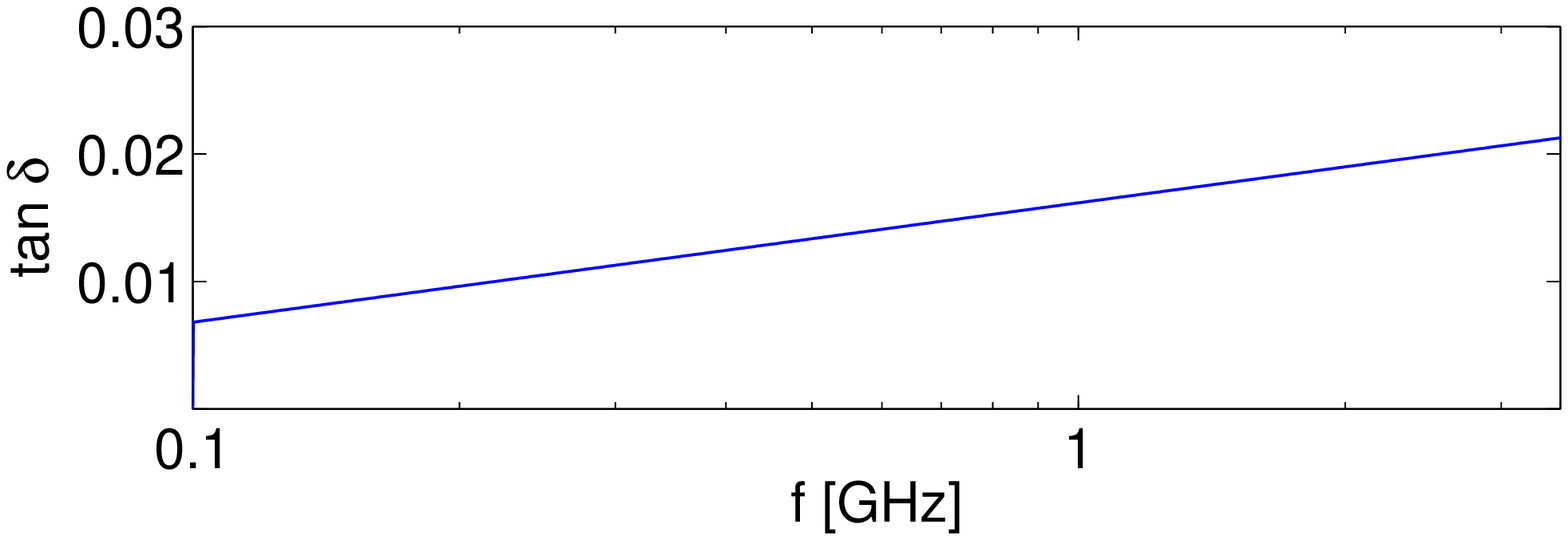}
\caption{Up: modulus of the transmission coefficient $S_{21}$ in a 2m-long strip placed above a 3mm-thick
glass stack both for measurements (green/gray) and simulations including losses (blue/dark).
The inter-peak distance provides a value for $\epsilon_r=5.5$ within 5\% in the
range f=[0.1-1]GHz (upper inset, up to 0.5GHz). The lower inset shows the same measurement along
50cm length (the same stack but with the strip rotated). Down: a functional description of the loss-tangent
that provides a good agreement with transmission measurements above 0.5GHz.}
\label{frec_1strip}
\end{figure}

 We characterized the float glass employed by measuring
its frequency response to transmission along a 2.5cm-wide strip. Measurements were
 done with a 4-port network analyzer (3.5GHz bandwidth) with the strip
placed along a 2m-long glass stack placed over a quasi-infinite ground plane (like in previous section),
but we performed a control measurement by
placing it transversally (0.50m length) and so reducing the losses.
The space between strip and ground was filled with three 1mm-thick glass plates, and pressure was applied via
polystyrene foam as described in previous section. Additional glass plates
were placed above for evaluating possible systematic errors; we also compared the results for a G10-supported
strip with the ones for standard Cu tape of the same width. In all cases the differences were minimal. Fig.
\ref{frec_1strip} shows the measured and simulated modulus of the transmission coefficient $S_{21}$
for a G10-supported strip and no additional glass plate. The
electrostatic parameters $C_g$ and $L_0$ were obtained from MAXWELL-2D. The best overall description
implies a value for $\epsilon_r=5.5$ for the glass,
that describes very well the inter-peak separation $\Delta f = v/(4\pi D)$ and the amplitude oscillations
at low frequencies (see insets). A close look at this observable allows to determine that the
dielectric constant is varying by 5\% at most in the range $f=[0.1-1]$GHz.
At higher frequencies, the observed behavior is dominated by the dielectric losses.

We tried to describe the transmission at high-frequencies by assuming a
constant value of the loss-tangent $\tan{\delta}=0.021$, that provides a reasonable description of the data. It
 over-estimates, however, the signal attenuation at low frequencies. The data favors a soft increase with
frequency in the range [0.1-3.5]GHz, that we have operationally parameterized with a logarithm
(table \ref{tan delta}), for simplicity:
\beq
\tan{\delta}(f[\tn{GHz}]) = \tan{\delta}_{0.1} + \frac{\tan{\delta}_{3}-\tan{\delta}_{0.1}}{\log_{10}{\frac{3}{0.1}}} \log_{10}\frac{f}{0.1} \label{tloss}
\eeq
Such a smooth increasing behavior at ambient temperature 
is qualitatively compatible with the one reported in \cite{Frech}.

\begin{table} [h!!]
\begin{center}
\begin{tabular}{|c|c|c|c|c|}   \hline
                          & $\tan{\delta}_{0.1}$ & $\tan{\delta}_{3}$ & $\tan{\delta}$ & $f_c$[GHz]\\ \hline
    D=2m, only glass      & 0.007 & 0.021 & 0.021 & 0.55\\ \hline
    D=0.5m, only glass    & 0.007 & 0.032 & 0.03  & 1.3\\ \hline
    D=2m, 2-gaps/3-glass  & 0.007 & 0.032 & 0.029 & 0.85\\ \hline
\end{tabular}
\caption {\footnotesize Table with the best description of the observed losses for three different structures.
The first two columns show the 2 parameters obtained by assuming a logarithmic increase with frequency
in the range $f=[0.1-3.5]$GHz (eq. \ref{tloss}). The third column shows the best description assuming a constant value. The
last column shows the cutoff frequency for each structure.}
\label{tan delta}
\end{center}
\end{table}

Apart from the measurements for a 3mm-thick 2m-long
stack, additional measurements on an 0.5m-long stack and a 2-gap/3-glass
structure were performed, and the best values obtained for the $\tan{\delta}$ of the glass are given
in table \ref{tan delta}. The third column shows the value for $\tan{\delta}$ when assumed to be constant
over the whole frequency range, from which an average value $\tan{\delta}=0.25\pm 0.05$ in the
range f=[0.1-3.5]GHz can be inferred for the float
glass we used, dominated by systematic uncertainties. The cutoff frequency $f_c$ is also given
in the last column.
Due to the highly oscillatory pattern,
it was determined from a comparison with the simulated $|S_{21}|$, as in Fig. \ref{frec_1strip}.
The value of $f_c$ was then obtained through evaluation of the condition $e^{-D/\Lambda}=1/\sqrt{2}$.

Note that the cutoff frequency is as small as $f_c=0.55$GHz for the propagation over
a 3mm-thick glass stack along 2m. The situation
improves when including gas gaps, up to $f_c=0.85$GHz, but still
far from the intrinsic cutoff frequency $f_c=3$GHz expected for RPC signals.
Transmission is indeed limited essentially by the dielectric losses in
the glass since the attenuation due to resistive losses is as small as $1/1.05$ at 3GHz.
As said, the dependence of $\Lambda_{_G}$ and $\Lambda_{_R}$
with the particular geometry is very small for wide-strip RPCs so skin effect will play usually
 a minor role in this type of detectors.

\subsection{Deviation from quasi-TEM propagation}

\begin{figure}[ht!!!]
\centering
\includegraphics*[width=\linewidth]{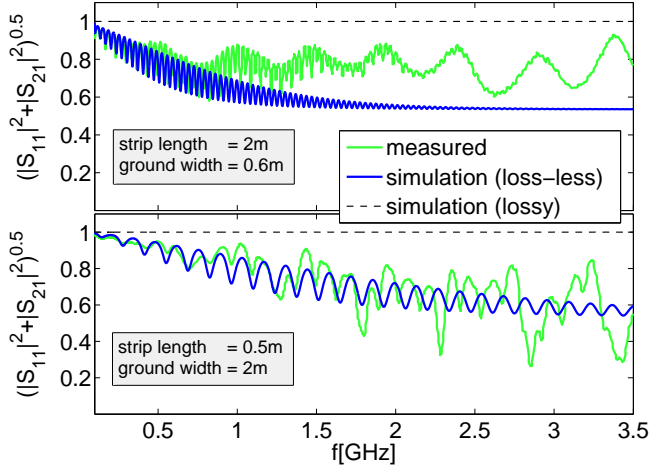}
\caption{Up: Sum-coefficient $S=\sqrt{|S_{21}|^2+|S_{11}|^2}$ for measurements (green/gray),
simulations assuming losses (blue/dark), and simulations without losses ($|S|=1$, dashed). Down: the same as
above but with the strip rotated (ground plane approximately $\times 4$ wider). Deviations from
a pure TEM description are suppressed in the lower case, due to the ground plane being closer to the ideal
infinite limit.}
\label{beyondTEM}
\end{figure}

The dumped oscillating behavior above 1GHz in Fig. \ref{frec_1strip}, overlaid on the pure
system losses can not be accommodated in a simple quasi-TEM image, and is indeed present for all measurements
on 2m-long/0.6m-wide structures. This
discrepancy can be highlighted by studying the sum-coefficient $S=\sqrt{|S_{21}|^2+|S_{11}|^2}$,
that is shown in Fig. \ref{beyondTEM}. Measurements (green/gray line), simulations assuming losses
(blue/dark line) and without losses ($|S|=1$, dashed) are plotted for two cases: in Fig. \ref{beyondTEM}-up the transmission
 is measured along 2m over an 0.6m-wide ground plane while in Fig. \ref{beyondTEM}-down the
  transmission is measured on the rotated structure, meaning an 0.5m-long strip over a 2m-wide ground plane,
the later showing a better agreement with simulations.

It must be noted that the critical frequency for the lowest TE (transverse electric) mode to start
propagating along a strip placed over an infinite ground plane at close distance from it is:
\beq
 f_{_{TE}}> \frac{v}{2\tn{w}} \label{f_TE}
\eeq
where $\tn{w}$ is the width of the strip, thus yielding $f_{_{TE}}>7$GHz. The frequency at which
simulations and measurements deviate in Fig. \ref{beyondTEM}-up is, however, $f_{_{TE}} \simeq 0.3$GHz,
 corresponding to a width of $\simeq60$cm,
 exactly as our ground plane. Incidentally, measurements in Fig. \ref{beyondTEM}-down over the rotated structure
 ($f_{_{TE}} \simeq 0.075$GHz)  show a much better agreement with a quasi-TEM description based uniquely
 on eqs. \ref{S21}, \ref{S11}. In order to better understand this fact, it must be
recalled that propagation of non-TEM modes is theoretically forbidden
below the cutoff frequency when the widths of both planes
 (strip and ground) satisfy simultaneously condition \ref{f_TE} (see \cite{Collins}, for instance).
Since an infinite ground plane mimics the presence of a mirror strip at equal distance from
it, the same condition applies for this latter case, as long as the plane is infinite in
extension. This would explain the better agreement
 with a TEM-description for a wider ground plane, as in Fig. \ref{beyondTEM}-down.
In any case, it will be shown in the next section that these deviations from pure quasi-TEM
propagation do not need to be included in order to accurately reproduce the typical figures of interest
in RPC transmission.

\subsection{Losses in 2-strip structures}

Unfortunately, the solutions to the Transmission Line problem including losses become
immediately non-analytic for more than 1 strip \cite{Sarkar}. Our proposal is to assume that the eigen-vectors
and eigen-values of the problem are very slightly modified in the presence of
losses, so that the structure of the solutions remains the one provided by
the loss-less problem, and losses can be included as a convolution at a later
stage. Additionally, the cross-conductance between strips is neglected. In order to
experimentally demonstrate this assertion, measurements
on 2-strip structures were performed, and simulations carried out
under this factorization assumption. We first need to adapt the solutions given by 
eq. \ref{full_blast} to the case where measurements are performed with a network analyzer.
For this we recall that the network analyzer was operated with standard terminations
($R=50\Omega$) and that eq. \ref{full_blast} can be easily adapted
to this particular case, similarly to previous section. The array of voltages measured at
the far-end in the time 
domain for a voltage source $V(t)$ (as measured on $R=50\Omega$) connected to strip $n$ 
can be then obtained as:
\beq
\vec{V}_{_{FE}}\!(t)\! =\! \hat{T}\! \sum_{j=0}^{\infty}(1\!-\!\hat{T})^j\!
\hat{M}\!\!\!\left(\!\!\begin{array}{c} M^{-1}_{1n}V(t\!-\!\frac{(\!-1\!)^j\! D \! + \! 2\lceil{j/2}\rceil\!{D}}{v_1})
\\\ldots\\ M^{-1}_{Nn}V(t\!-\!\frac{(\!-1\!)^j\!D \! + \! 2\lceil{j/2}\rceil\!{D}}{v_N})  \end{array}\!\!\right) \label{full_blast_St}
\eeq
That is nothing else but eq. \ref{full_blast_2} for N-strips under the equivalence \ref{ref_I}.
The calculation of the voltages at the near-end $\vec{V}_{_{NE}}(t)$ is analogous, but requires
the incoming voltage pulse to be explicitely considered at the given port: 
\bear
&&\vec{V}_{_{NE}}\!(t)\! = \hat{T}\! \sum_{j=0}^{\infty}(1\!-\!\hat{T})^j\!
\hat{M}\times \nonumber\\
&&\times \left(\!\!\begin{array}{c} M^{-1}_{1n}V(t\!-\!\frac{D-(\!-1\!)^j\! D \! + \! 2\lfloor{j/2}\rfloor\!{D}}{v_1})
\\\ldots\\ M^{-1}_{Nn}V(t\!-\!\frac{D-(\!-1\!)^j\!D \! + \! 2\lfloor{j/2}\rfloor\!{D}}{v_N})  \end{array}\!\!\right) 
- \left(\begin{array}{c} \ldots \\ 0_{n-1} \\ V(t) \\ 0_{n+1} \\ \ldots \end{array}\right) \label{full_blast_Sr}
\eear
The scattering matrix parameters can be obtained by directly 
computing the transmission at different frequency components through a Fourier transform:
\beq
\vec{S}_{_{FE,NE}}(f)=\frac{\tn{ft}(\vec{V}_{_{FE,NE}}(t))}{\tn{ft}(V(t))} \label{FT1}
\eeq
This equation provides, indeed, the solutions to the loss-less situation. Our proposal for
including losses is the ansatz:
\beq
\vec{S}_{_{FE,NE}}(f)|_{lossy}
= \vec{S}_{_{FE,NE}}(f)|_{loss-less} \times \exp(-\frac{D}{\Lambda(f)}) \label{factoriza}
\eeq
that, translated into the time-domain, effectively implies a convolution of the solutions in
eq. \ref{full_blast}, \ref{full_blast_St} or  \ref{full_blast_Sr}
with the inverse Fourier transform of $\exp(-\frac{D}{\Lambda(f)})$. As can be
deduced from the functional dependence of $S_{11}$  (eq. \ref{S11}) this
ansatz is not accurate for the reflection coefficient. This is fortunately not
a relevant parameter to understand transmission of signals induced inside
a counter, so it is not considered further. Thus, by using the port numbering scheme in  Fig. \ref{scheme}:
\beq
S_{2,3,41}(f)|_{lossy}= S_{2,3,41}(f)|_{loss-less} \times \exp(-\frac{D}{\Lambda(f)}) \label{factoriza2}
\eeq
\begin{figure}[ht!!!]
\centering
\includegraphics*[width=\linewidth]{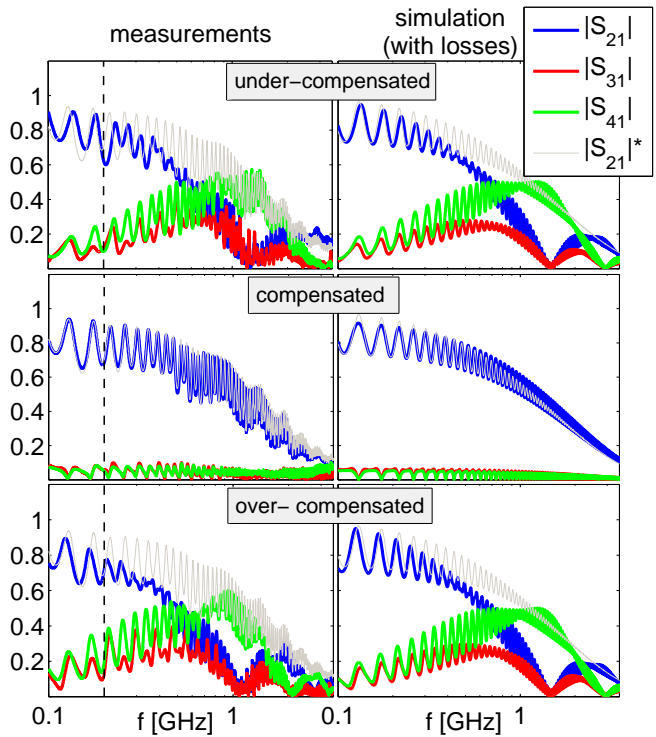}
\caption{Left: the measured moduli of the transmission
coefficient $S_{21}$, far-end cross-talk $S_{41}$ and near-end cross-talk
$S_{31}$ in a 2-gap/3-glass structure, with 2 strips separated by 3.1mm. From up
to down, an under-compensated case (no additional glass),
compensated (one additional glass plate) and over-compensated (three additional
glass plates). $S_{21}^*$ is the transmission coefficient for 1-strip in the,
otherwise, same structure. Right: simulations for the general lossy case
under the factorization ansatz proposed in text. The vertical line shows
the range previously measured in \cite{Riegler_transm}.}
\label{frec_all}
\end{figure}
In order to numerically obtain the scattering matrix parameters we have assumed an initial voltage $V(t)$
with fast enough components in the range studied here. For simplicity, an exponential signal
with $100$ps rise-time has been used. The procedure was:
\begin{enumerate}
\item Obtain the analytical solutions to the problem (eqs. \ref{full_blast_St}, \ref{full_blast_Sr}).
\item Make the Fourier transform, according to eq. \ref{FT1}.
\item Apply the attenuation factor $\exp(-\frac{D}{\Lambda(f)})$ obtained from simulations of
the 1-strip transmission coefficient $S_{21}$ as in Fig. \ref{frec_1strip}.
\end{enumerate}

Fig. \ref{frec_all} shows the measurements of the scattering matrix coefficients (left)
and the corresponding simulations (right) for the 2-gap/3-glass
structure previously studied in Fig. \ref{waveforms} in a time-domain representation.
The 1-strip transmission coefficient is shown as $S_{21}^*$(light grey)
for reference.
From up-down three cases are presented: under-compensated (no additional glass plate),
compensated (one additional glass plate) and over-compensated (three additional glass plates).
We note that the observed frequency pattern of
the compensated system is strikingly simple, as predicted in Fig. \ref{frec_res}, allowing
for an extended bandwidth (up to the limit imposed by losses) and much reduced cross-talk
patterns.

\begin{figure}[ht!!!]
\centering
\includegraphics*[width=8.5 cm]{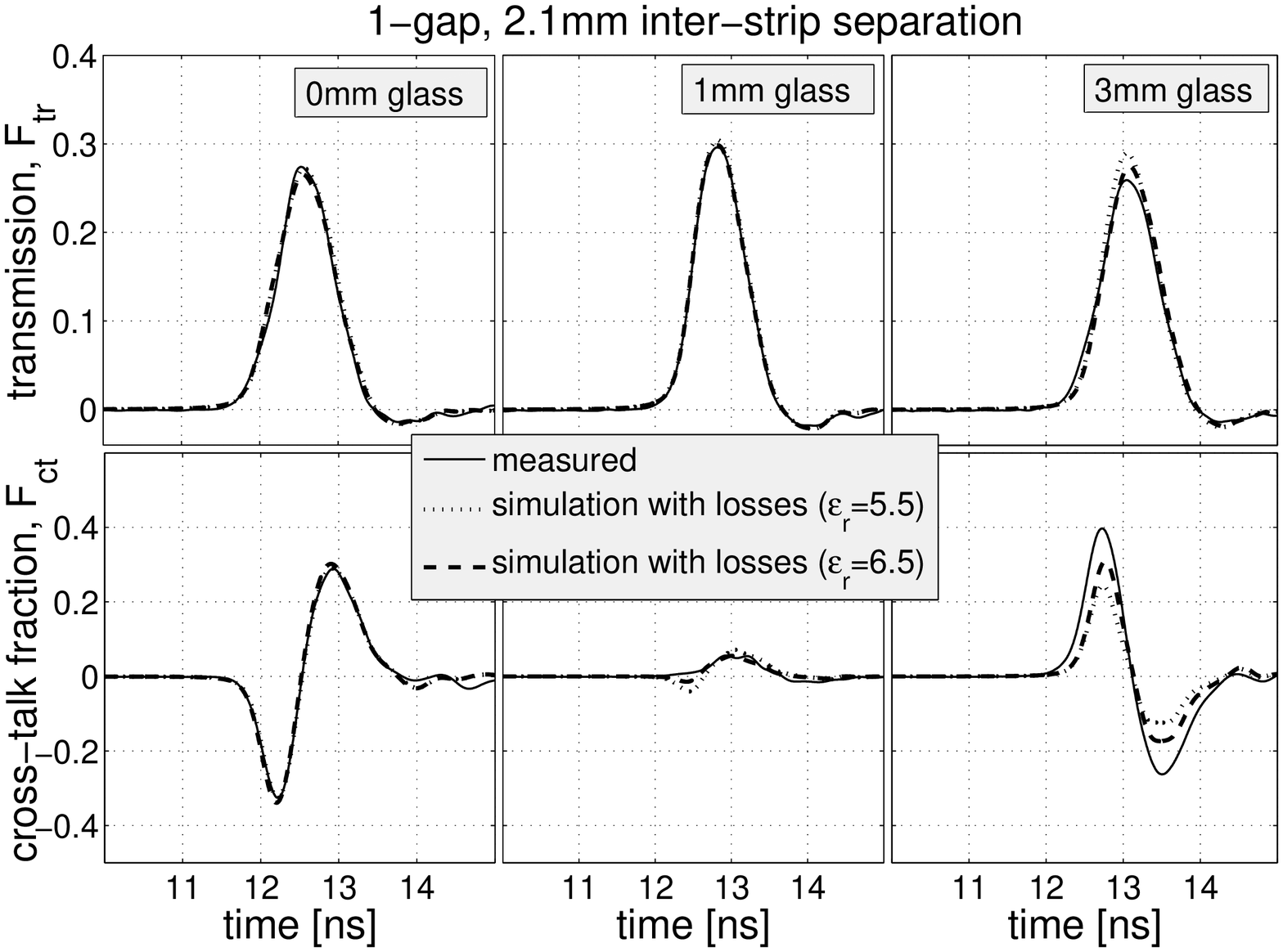}

\includegraphics*[width=8.5 cm]{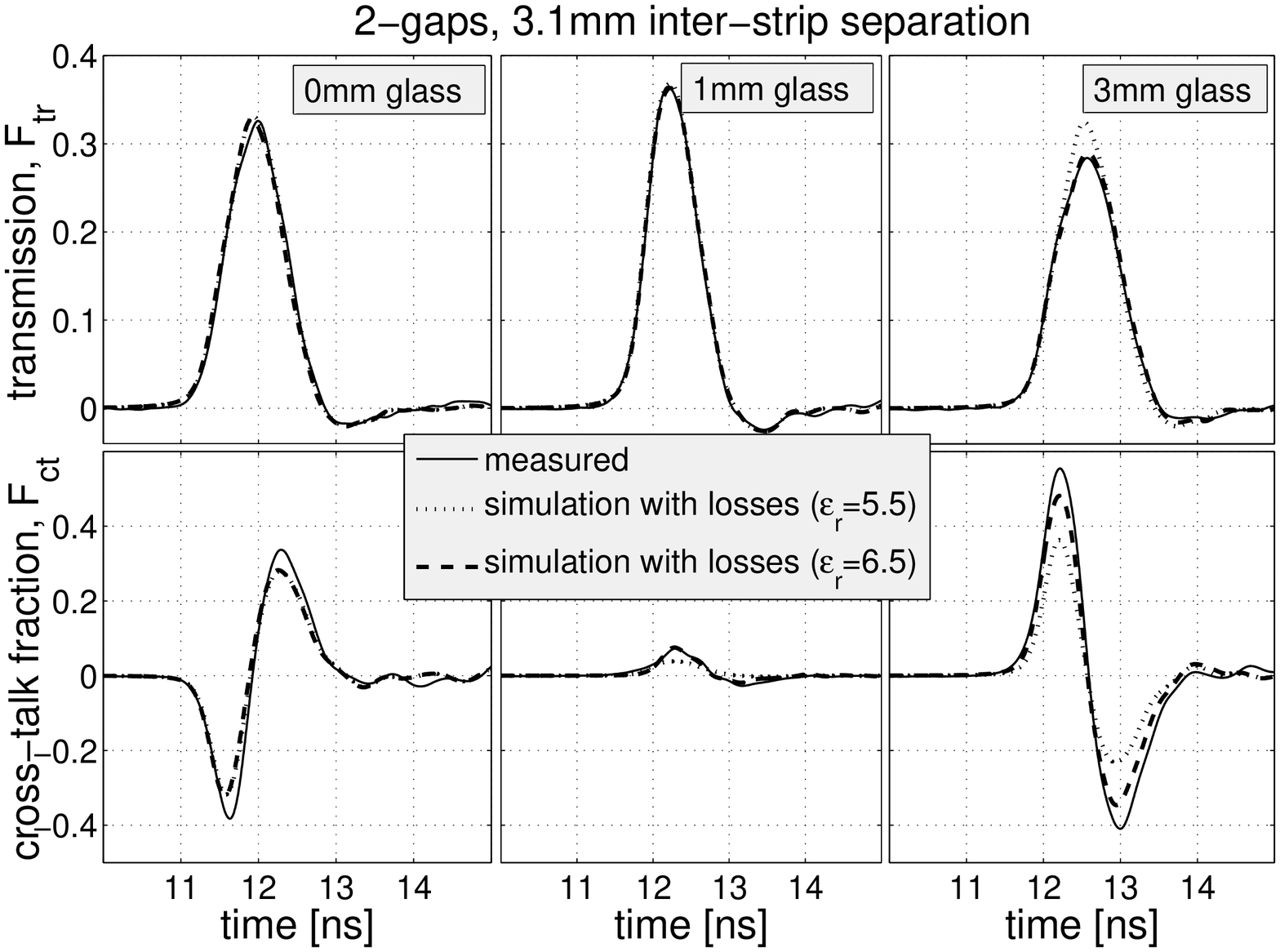}
\caption{Detailed comparison between full simulations including losses (dashed, dot-dashed)
and the measured oscillograms (continuous lines). 
The comparison is made in the reflection-free region for the 1-gap RPC structure (up)
and the 2-gap one (down) previously discussed in Fig. \ref{waveforms}. 
From left to right, the cases `under-compensated'
(no additional glass), `compensated' (1mm additional glass) and `over-compensated'
(3mm additional glass) are shown. Although a parameter-free simulation based on the directly measured
value of $\epsilon_r=5.5$ provides a reasonable agreement, a value $\epsilon_r=6.5$
gives the best overall description.}
 \label{waveforms_zoom_all}
 \end{figure}

\begin{figure}[ht!!!]
\centering
\includegraphics*[width=9.5 cm]{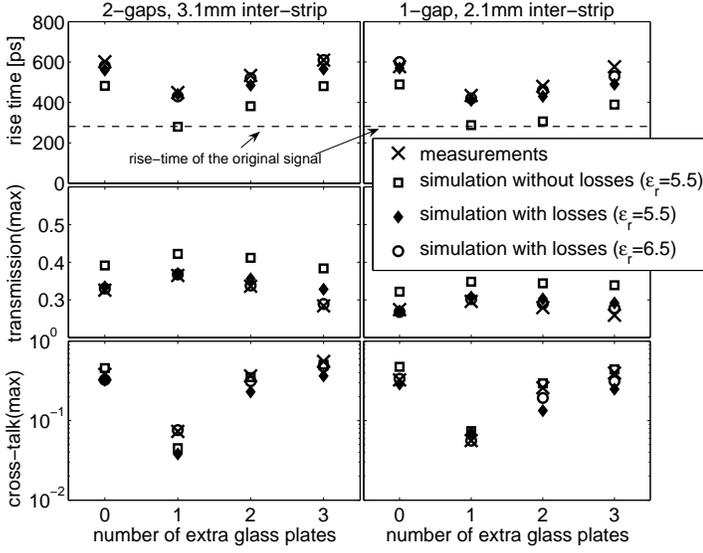}
\caption{Comparison between measurements and simulations for
the main characteristics of a typical signal (signal rise-time,
maximum of the signal transmission and cross-talk fraction). The signal characteristics
are very well described after including losses, where the value
$\epsilon_r=6.5$ for glass can be considered as an optimum choice.}
\label{exp_meas_with_losses}
\end{figure}

In order to reconcile the measurements in frequency domain in Fig. \ref{frec_all} with the ones in time-domain
in Fig. \ref{waveforms} we have convoluted (under the factorization assumption  \ref{factoriza2}) the time-domain
simulated waveforms in previous section with
the inverse of the Fourier transform of $\exp(-\frac{D}{\Lambda(f)})$. The results
are shown in Fig. \ref{waveforms_zoom_all} by zooming-in the direct signal (no reflections).
Time-offsets have been adjusted in order to allow precise shape-comparisons, for which
the simulations with $\epsilon_r=5.5$ (dotted) are used as reference, and both
measurements (continuous) and simulations with $\epsilon_r=6.5$ (dashed) are time-shifted
to give the best possible agreement. A detailed comparison of the most relevant
observables is given in Fig. \ref{exp_meas_with_losses}.

Despite the overall good agreement, simulations clearly favor a value for $\epsilon_r=6.5$, contrary
to the measurements in the frequency domain, that favor $\epsilon_r=5.5$. Of course,
by using  $\epsilon_r=6.5$ the propagation velocity, that is directly
connected to the inter-peak distance in the frequency domain,
(and is not included in Fig. \ref{exp_meas_with_losses})
is not properly described in the time-domain representation, by some 4\% percent.
This discrepancy is presumably stemming from plate to plate variations or
slight material anisotropies, and it has
to be understood as the systematic uncertainty present in these set of measurements.

\section{Electrostatic compensation in N-strip structures}\label{section4}

A literal solution for N-strips,
like the one given in \cite{Diego} for 2-strips, could be useful for counter optimization.
Besides, it is important to understand to which extent the ideas introduced in previous
section for the simple 2-strip case can be applied to an N-strip situation.
Inspired by the work of Romeo and Santomauro \cite{Santo},
we propose to approximate $\hat{C}$, $\hat{C}_0$
by tri-diagonal matrices. Translational invariance across the strips is
also assumed, so that the electrostatic couplings of each strip are the same
with respect to its neighbors and ground.

For simplicity we study the 5-strip case that, we believe, contains
the essential features of an N-strip RPC system. Thus, our input matrices have the form:
\beq
\hat{X}=\left(\begin{array}{ccccc} X_0 & -X_m & 0 & 0 & 0\\ -X_m & X_0& -X_m & 0 & 0\\
0 & -X_m & X_0 & -X_m & 0\\0 & 0 & -X_m &X_0 & -X_m\\0 & 0 & 0 & -X_m & X_0\\ \end{array}\right) \label{3diag}
\eeq
where $\hat{X}$ stands for either $\hat{C}$ or $\hat{C}_0$. In \cite{Santo} indeed, the assumption
was made that $\hat{L}$ itself was tri-diagonal instead of $\hat{C}_0$. It seems to us
 that the latter approach provides generally a higher accuracy for RPC structures and, besides, allows
 for slightly more compact analytical formulas; we will refer to this in the following
 as the `short-range coupling' approximation. The structures of previous section,
for instance, show a value for the coupling coefficient $C_m/C_0\sim1/10$,
while the (eventual) coupling to a second neighbor, in a translationally invariant structure,
would be $C_{m2}/C_0\lesssim1/100$, so that
assumption \ref{3diag} is a priori justified.\footnote{Note that $C_0=C_g+C_m$ in the 2-strip case.
In general, it is not easy to strongly violate any of these
two conditions for typical RPC designs. A remarkable exception is \cite{FOPI}, due the fact
of $\tn{w}/h \ll 1$.} Beyond the 2-strip case, the `short-range coupling' approximation implies
an artificially low coupling to ground for the first and the last strips (at the sides of the
structure). This is so because the $\hat{C}$-matrices are defined such that the capacitance
to ground of strip $i$ satisfies the condition
$C_{g,i} = \hat{C}_{ii} - \sum_{j=1,j\neq{i}}^N \hat{C}_{ij}$, ${C}_{ij}$ are defined negative,
and those strips have one neighbor less. Signals produced at the
boundary-strips of the structure will, therefore, 
show a deviation from the behavior predicted by this approximation. 
Their overall influence in propagation will be increasingly small as the number
of strips increases.
The exact solution to the diagonalization problem \ref{dia_pro} under the `short-range coupling'
approximation
(taking for commodity of operation the unitary representation) has a simple form independently
from the particular values of the elements of the $\hat{C}$-matrices:
\beq
\hat{M}=\hat{M}^{-1}=\frac{1}{2}\left(\begin{array}{ccccc} \frac{1}{\sqrt{3}} & -1 &\frac{2}{\sqrt{3}} & -1 & \frac{1}{\sqrt{3}}
\\ -1 & 1 & 0 & -1 & 1\\\frac{2}{\sqrt{3}} & 0 & -\frac{2}{\sqrt{3}} & 0 & \frac{2}{\sqrt{3}}
\\-1 & -1 & 0 & 1 & 1
\\\frac{1}{\sqrt{3}} & 1 & \frac{2}{\sqrt{3}} & 1 & \frac{1}{\sqrt{3}}\\ \end{array}\right) \label{M_mat}
\eeq
and the components of $\vec{v}$ are:
\bear
\vec{v}= && c \Bigg\{\sqrt{\frac{C_{00}+\sqrt{3}C_{m0}}{C_0+\sqrt{3}C_m}},  \sqrt{\frac{C_{00}+C_{m0}}{C_0+C_m}},
\sqrt{\frac{C_{00}}{C_0}}, \nonumber \\
 && \sqrt{\frac{C_{00}-C_{m0}}{C_0-C_m}}, \sqrt{\frac{C_{00}-\sqrt{3}C_{m0}}{C_0-\sqrt{3}C_m}}\Bigg\} \label{v_vec}
\eear
The above velocity spectra becomes fully degenerated when the velocity dispersion,
defined as eq. \ref{par1}-right, equals zero and thus:
\beq
\frac{C_m}{C_{0}}=\frac{C_{m0}}{C_{00}} \label{comp5}
\eeq
 meaning that modal dispersion is exactly canceled, and all modes travel at a velocity given by $\bar{v}$
 in eq. \ref{par1}-left. Indeed, this crucial property
can be easily generalized to an arbitrary number of strips. For this it is necessary to realize that the condition
\ref{comp5} implies, under the `short-range coupling' approximation, that the matrices $\hat{C}$ and $\hat{C}_0$
differ in a multiplicative constant. Thus, the matrix to be diagonalized in the basic problem of eq. \ref{dia_pro}
becomes, after recalling eq. \ref{LvsC}, proportional to the unit-matrix, whose eigenvalues are
obviously identical. This result means, precisely, that any inhomogeneous N-strip transmission line
that is translationally invariant across the strips, does not
consist of magnetic materials and falls under the short-range approximation will show no modal dispersion
when it is electrostatically compensated. Interestingly, this will hold even if the coupling
to the $1^{st}$ neighbor is high (low coupling approximation not fulfilled).

Despite the simplicity of the solution to the diagonalization problem in eqs. \ref{M_mat}, \ref{v_vec}, a
tractable literal solution requires of further
assumptions due to the need to obtain a simple expression for the transmission coefficient $\hat{T}$
in eq. \ref{full_blast}.
We propose the `low coupling' assumption:
\beq
\frac{C_m}{C_0}, \frac{C_{m0}}{C_{00}}< 1 \label{low_cop}
\eeq
that is familiar from the 2-strip situation \cite{Diego}, where it has been defined through
the equivalent condition $Z_m/Z_c < 1$. It is also assumed `moderate dispersion', that means
that eq. \ref{comp5} is always approximately satisfied. Under this latter assumption we keep the
lowest order dispersive terms, that are the ones proportional to the difference of the traveling modes. At last,
we inject a signal $I(t)$ at position $y_0$ along the second strip and obtain the transmitted and cross-talk
signals in the $3^{rd}$ ($1^{st}$ neighbor) and $4^{th}$ ($2^{nd}$ neighbor) strips, in order to avoid side effects.

The literal solutions, omitting reflections, can be finally obtained:

\begin{eqnarray}
I_{tr}(t)   \simeq && \frac{T}{2}\frac{\Sigma_1 I (t,y_0) + \Sigma_2 I (t,y_0)}{2} +{}  \nonumber \\
&& \!\frac{Z_m R}{(Z_c+R)^2} \frac{\sqrt{3}\Delta_1 I (t,y_0) + \Delta_2 I (t,y_0)}{2} \label{tr_5}\\
I_{ct,1}(t)   \simeq && \frac{Z_m R}{(Z_c+R)^2} \Sigma_1 I (t,y_0)\! +\! \frac{T}{2\sqrt{3}} \Delta_1 I(t,y_0) \label{ct1_5} \\
I_{ct,2}(t)   \simeq &&\frac{(Z_m R)^2}{(Z_c+R)^3} \frac{1}{Z_c} \frac{\Sigma_1 I (t,y_0) + \Sigma_2 I (t,y_0)}{2} + \nonumber \\
&&\frac{T}{2} \frac{\Sigma_1 I (t,y_0) - \Sigma_2 I (t,y_0)}{2} \label{ct2_5}
\end{eqnarray}
with:
\begin{eqnarray}
&& \Sigma_{1(2)}{I}(t,y_0) = \frac{1}{2}\! \left[ I(t-\frac{y_0}{v_{5(4)}}) + I(t-\frac{y_0}{v_{1(2)}}) \right] \label{mode1} \\
&& \Delta_{1(2)}{I}(t,y_0) = \frac{1}{2}\! \left[ I(t-\frac{y_0}{v_{5(4)}}) - I(t-\frac{y_0}{v_{1(2)}}) \right] \label{mode2}
\end{eqnarray}
The structure of the solutions in eqs. \ref{tr_5}-\ref{ct2_5} highlights the formal analogy with the
literal solutions to
the 2-strip problem given in \cite{Diego}. When the system
is compensated and reflections are neglected, transmission becomes
independent from the signal shape, the propagation distance and the position and thus:
\begin{eqnarray}
&& F_{tr}      = \frac{T}{2} \\
&& F_{ct,1}    = \bigg[\frac{R}{Z_c+R}\bigg] \frac{Z_m}{Z_c} \\
&& F_{ct,2}    = \left( \bigg[\frac{R}{Z_c+R}\bigg] \frac{Z_m}{Z_c} \right)^2 = F_{ct,1}^2 \label{ct2_comp}
\end{eqnarray}
implying a flat time (and frequency) response.

\begin{figure}[ht!!!]
\centering
\includegraphics*[width=\linewidth]{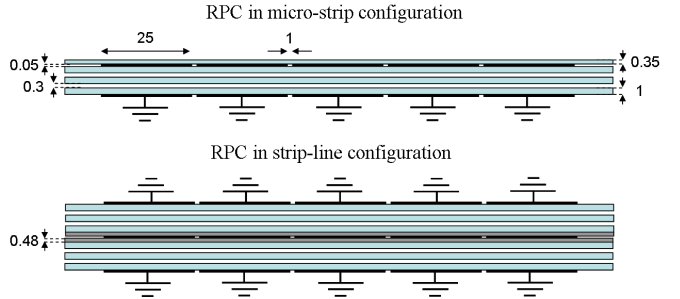}
\caption{Two electrostatically compensated 5-strip RPCs studied in this section.}
\label{5strip_fig}
\end{figure}

In order to illustrate the power of
the electrostatic compensation technique we have performed simulations for two realistic
5-strip RPCs, each strip being 2.5cm-wide. One of the structures is a 2-gap/3-glass micro-strip
structure (like in previous sections), but the ground plane has been also segmented and
consists of 5 strips identical to the signal strips. The second structure is a strip-line
structure. Precisely, it is the same micro-strip structure but mirrored with respect
to the signal plane. The inter-strip separation is as small as 1mm in both cases. Compensation has been
achieved in the first case by adding 0.35mm thick glass over the signal plane ($\epsilon_r=5.5$),
while in the second case 2 additional layers of Teflon ($\epsilon_r=2.1$), 0.48mm-thick, have
been added up and down with respect to the signal plane (Fig. \ref{5strip_fig}). A compilation
of the main elements of the capacitance matrices is given in table \ref{C-matrix}.

\begin{table} [h!!]
\begin{center}
\begin{tabular}{|c|c|c|c|c|c|c|}   \hline
    structure                   & $C_0$ & $C_m$ & $C_{m2}$ & $C_{00}$ & $C_{m0}$ & $C_{m02}$ \\ \hline
    micro-strip (uncomp.) & $276.2$ & $36.1$  & $1.23$  & $98.7$ & $15.7$ & $1.21$ \\ \hline
    micro-strip (comp.)   & $299.7$ & $47.8$  & $1.26$  & $98.7$ & $15.7$ & $1.21$ \\ \hline
    strip-line (uncomp.)  & $499.8$   & $49.6$  & $0.0006$  & $144.9$ & $8.9$ & $0.00035$ \\ \hline
    strip-line (comp.)    & $390.0$ & $28.3$  & $0.00036$ & $131.4$ & $9.5$ & $0.00026$ \\ \hline
\end{tabular}
\caption {\footnotesize Compilation of the main capacitive elements, in pF/m, for various
5-strip RPCs under the following definitions
$C_0\equiv \hat{C}_{33}$, $C_{m}\equiv -\hat{C}_{23}$, $C_{m2}\equiv -\hat{C}_{13}$,
and similarly for the elements of matrix $\hat{C_0}$.}
\label{C-matrix}
\end{center}
\end{table}

\begin{figure}[ht!!!]
\centering
\includegraphics*[width=\linewidth]{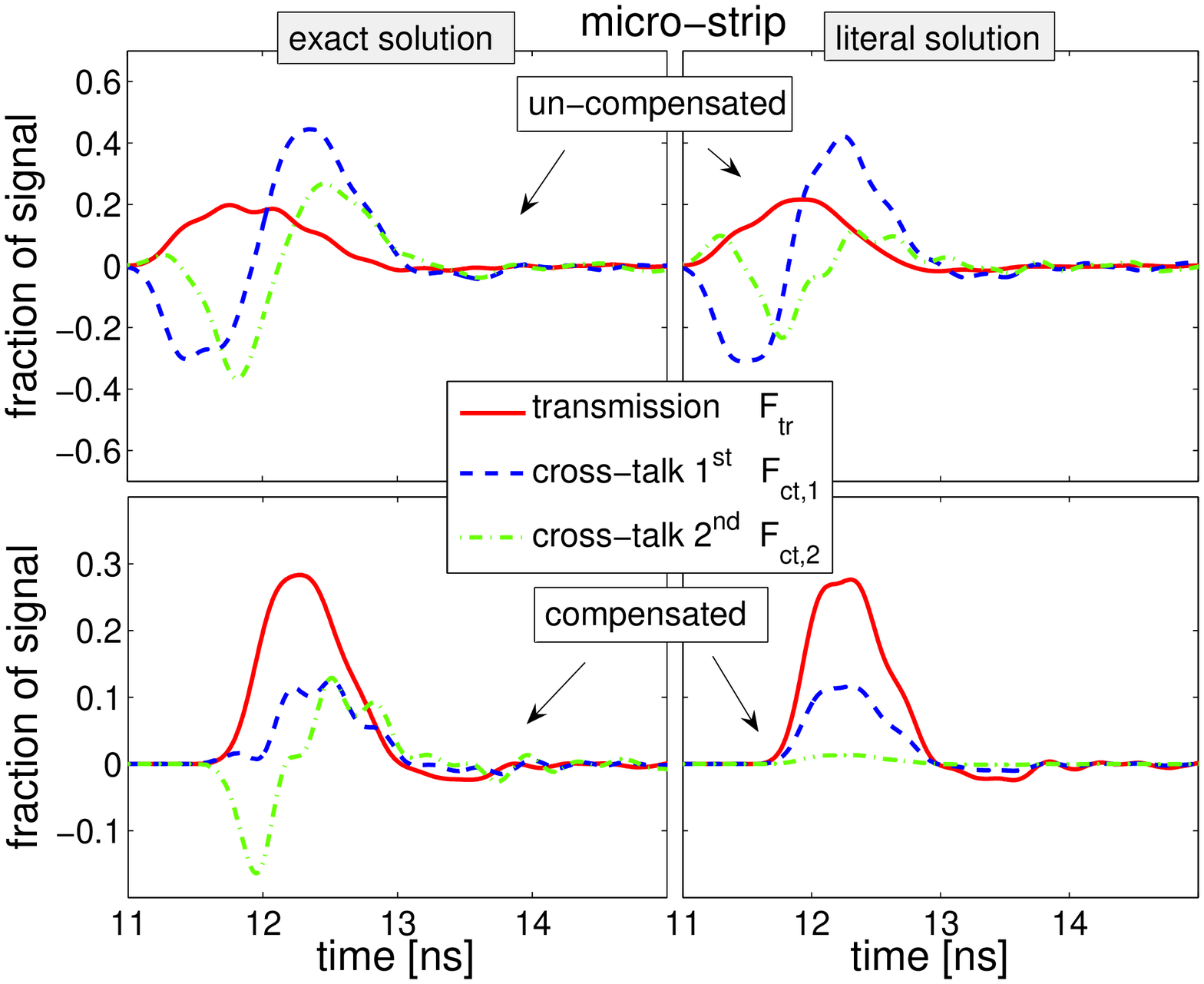}
\includegraphics*[width=\linewidth]{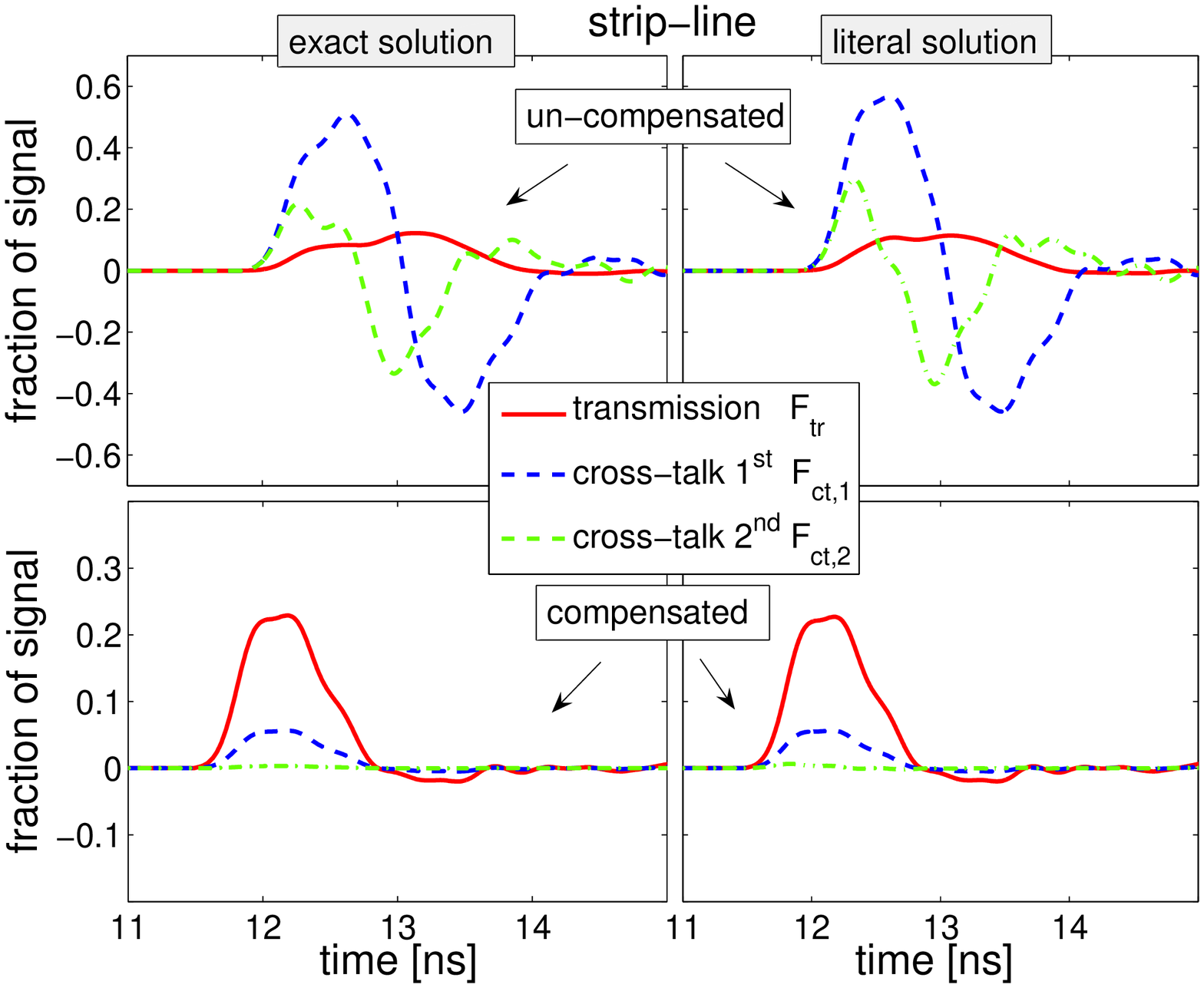}
\caption{Up-set: Simulations for transmission along 2m-long 5-strip RPC in a micro-strip configuration
as described in text, with 2.5cm-wide strips placed as close as 1mm. The left column shows the
exact solution and the right one the literal solutions given in text. The upper row stands for
a non-compensated case and the lower one for a compensated one. The fraction of
transmitted signal (red, continuous), fraction of cross-talk to the
first neighbor (blue, dashed) and fraction of cross-talk to the second neighbor
(green, dot-dashed) are shown. Low-set: like the upper set but for a 5-strip strip-line
configuration as described in text, also with 2.5cm width and 1mm inter-strip
separation.}
\label{5strip_all}
\end{figure}

Figs. \ref{5strip_all} show the results of the proposed optimization on 5-strip structures,
serving also as a comparison between the exact solutions (left) and
the literal solutions given in eqs. \ref{tr_5}-\ref{ct2_5} (right). The
same signal used in the measurements of previous section has been injected at the end
of a $2$m-long counter, and reflections have been omitted. Compensation has
been `optically' achieved in both micro-strip and
strip-line configurations, although the cross-talk level stays higher in the micro-strip case than
in the strip-line one. This is because of the presence of two close-by ground planes
in the strip-line configuration, unlike the micro-strip, thus confining the mutual coupling field to a
smaller volume in the former case and reducing the coupling coefficient (table \ref{C-matrix}).
Besides, the coupling to the second neighbor is almost nonexistent in this strip-line
structure, making the  `short-range coupling' approximation virtually exact, unlike the
micro-strip. Because of that,
compensation is exactly achieved, and the cross-talk to the second neighbor is as small
as $F_{ct,2} = F_{ct,1}^2\simeq 0.05^2$, from eq. \ref{ct2_comp}. Due to the higher
coupling to the second neighbor, this situation is not reached in the micro-strip
configuration, showing $F_{ct,2} \sim F_{ct,1}$.

\section{Discussion}\label{section5}

The measurements performed in sections \ref{section3} and \ref{section4} are equivalent
within the limits of the precision of the corresponding apparatuses. This equivalence has been
here presented by using the measured frequency response in Fig. \ref{frec_all} to simulate
the time-domain measurements in Figs. \ref{waveforms_zoom_all}, \ref{exp_meas_with_losses}.
Similarly, it is possible to use the Fourier-transformed
time-domain measurements in section \ref{section3} to describe the measurements performed
with the network analyzer in section \ref{section4}. This has been done and shows naturally
a good agreement, but with a much lower precision, specially above 1.5GHz, where the
frequency components of the pulser used (280ps rise-time) are much suppressed, and sensitivity
is lost. In particular,
this approach does not allow to isolate neither the losses nor the deviations from a TEM description
as precisely as it has been done with the network analyzer data. Besides,
Fourier-Transformed time-domain data has a much lower frequency sampling as compared with
the measurements from the network analyzer.

Despite we have found evidence of systematic variations of
both the glass loss-tangent $\tan\delta|_{glass}=0.25\pm0.05$ and its dielectric constant
$\epsilon_r=6\pm0.5$, a very satisfactory description of several data
taken in different conditions has been obtained. Additionally, `electrostatic compensation' has been
experimentally demonstrated for 1 and 2-gap structures allowing to obtain a factor
$\times 12$ reduction in cross-talk for a typical signal as measured with a $\sim 2$GHz
bandwidth system.
Thus, as it seems, the
frequency-dependence of the parameters, eventual anisotropies, as well as plate to plate variations and/or
line non-uniformities, in general, do not represent an obvious limitation when trying to
compensate RPC structures. Moreover,
an `a priori' design seems possible, if tight mechanical tolerances at the 0.1mm scale
are observed.

Transmission along the compensated systems here developed was limited only by the dielectric losses.
As shown, however, losses depend little on the system geometry for wide structures,
so there is no much room for optimization.
The most obvious (and maybe only) way consists on reducing the glass thickness or adding an
additional material with a sizable series capacitance and low loss-tangent (Teflon, for instance),
that imply a reduction of $\tan \delta^*$ through the factor $F$. If a proper selection
of the glass is included, it seems likely that the bandwidth of a compensated RPC based on float
glass could be extended virtually to the limit imposed by the intrinsic signal rise-times (3GHz)
up to 2m.

It must be noted that this is the first experimental verification of the existence
of modal dispersion in RPC structures, besides the indirect results in \cite{Diego}.
According to our simulations, the band-stop region
stemming from modal dispersion is centered around 600 MHz for the structures
previously studied in \cite{Riegler_transm}, far from the limit of the device used
for those measurements (200 MHz).

There are classical ideas on cross-talk minimization that have been also studied prior to the
elaboration of this work and will be here described generally. A final quantitative evaluation is
much system-dependent so only general arguments are given:
\begin{enumerate}
\item A ground plane or a cage close to the strip can arbitrarily reduce the coupling coefficient $Z_m/Z_c$,
thus will always reduce cross-talk for short strips, at the expense of a reduced impedance and
higher noise. However, compensation is not guaranteed by this procedure, so low cross-talk and high
transmission can not be ensured for long counters by just observing this principle.

\item Shielding vias or the usage of additional
materials with low dielectric constants can reduce the coupling coefficient
$Z_m/Z_c$. They take, however, a very small portion of the available field lines, that propagate
through the glass/gap structure anyhow. Modest improvements in transmission and cross-talk
can be nevertheless achieved.

\item We have presented here single-ended signal measurements on single-ended architectures.
On the other hand, differential signaling is known to give, in general, lower levels of cross-talk as compared to
single-ended signaling, and a faster decrease with the inter-trace separation \cite{diff}. This is
natural for large distances between each differential pair,
since the coupling becomes then identical for both traces of the same pair.
This is not, however, a typical situation for RPC
structures and we have found so far no evidence that this principle prevails.
It must be noted, however, that the signal induction process is indeed genuinely quasi-differential
(signals with opposite polarities and similar amplitudes propagate simultaneously in anode and
cathode of each strip). A comparison between differential signal injection (with ground strips let
floating and terminated by ideal ground)
and single-ended injection (where ground strips are assumed to be ideal ground) was performed
in \cite{Diego}, showing a negligible difference. These subjects clearly
deserve, nevertheless, further investigations.
\end{enumerate}

\section{Conclusions}\label{section6}

We say that a strip counter is `electrostatically compensated' when the coupling coefficient
$Z_m/Z_c \simeq C_m/C_0$ is the same in the filled and in the empty structure.
Equivalently, the capacitive and inductive coupling are balanced: $C_m/C_0=L_m/L_0$
and all the system velocities are equal. Propagation along a compensated structure
shows minimal cross-talk and dispersion patterns and it is limited only by losses. This
prediction has been experimentally demonstrated for 2m-long 2-strip RPCs,
allowing for a factor $\times 12$ cross-talk reduction at $\simeq 2$GHz bandwidth.
Losses have been experimentally isolated for the first time on glass RPCs, resulting in a typical
cutoff frequency of $f_c=0.85$GHz over 2m and a loss-tangent $\tan\delta=0.025 \pm 0.005$
for float glass. A prescription on how to theoretically include losses has been given and
compared with data, showing good agreement.

It has been shown that a condition sufficient for compensation to apply to any N-strip
structure is that the coupling is `short-range', meaning that the coupling to the second neighbor can be
neglected. Several realistic configurations, most prominently a wide strip-line,
satisfy this condition to a very large extent.

We expect that the literal solutions derived for the transmission and cross-talk
to first and second neighbors can be useful for fast (albeit approximate)
estimates of multi-strip transmission properties.

We have already started a systematic program aimed at implementing these ideas in large-area
multi-strip counters.

\ack
DGD is supported by HIC for FAIR and NSFC (project 11050110111/A050506), CH was supported by
BMBF and NSFC (project 11050110111/A050506). The authors want to acknowledge the
GSI staff H. Flemming, H. Deppe, P. Moritz, M. Ciobanu, M. Freimuth and P. Kowina for discussions
on this fascinating field and providing the necessary equipment, and to J. Hehner
for his always competent technical help. Professional assistance
with the matrix formalism from T. Rodr\' \i guez Frutos and K. Vartortout (also GSI) is acknowledged.
 N. Herrmann deserves credit for being the first to spot the improbable possibility of compensating
 modal dispersion.

\appendix

\section{Charge conservation of the solutions to the loss-less 2-strip problem}

The general solution to the 2-strip loss-less problem for a signal injected in strip
1 when termination is performed with individual resistors of equal value is:
\beq
\vec{I}_{T}\!(t)\!\! = \!\frac{\hat{T}}{2} \!\! \sum_{j=0}^{\infty}\! (1\!-\!\hat{T})^j\!
\hat{M}\!\!\left(\!\!\begin{array}{c} \hat{M}^{-1}_{11}\!I(t\!-\!\frac{(\!-1\!)^j y_0 \! + \! 2\lceil{j/2}\rceil\!{D}}{v_1})
\\ \hat{M}^{-1}_{12}\!I(t\!-\!\frac{(\!-1\!)^j y_0 \! + \! 2\lceil{j/2}\!\rceil{D}}{v_2})  \end{array}\!\!\right) \label{full_blast_app}
\eeq
that was given in \cite{Diego}.\footnote{There is indeed an erratum in formula (12) in ref \cite{Diego},
the matrix $\hat{T}$ should appear before the sum, as in eq. \ref{full_blast_app}.} The charge finally
transmitted through each line for a signal injected in strip 1 can be obtained after
integration of eqs. \ref{full_blast_app} over infinite time, yielding:
\beq
\vec{q}_{T} = \frac{\hat{T}}{2} \sum_{j=0}^{\infty} (1-\hat{T})^j
\hat{M}\left(\begin{array}{c} \hat{M}^{-1}_{11}q
\\ \hat{M}^{-1}_{12}q \end{array}\right) \label{full_blast_q}
\eeq
where $q$ is the charge conveyed by the induced signal $I(t)$. For simplicity we take the unitary representation:
\beq
\hat{M}=\hat{M}^{-1}=\frac{1}{\sqrt{2}}\left(\begin{array}{cc} 1 & 1 \\ 1 & -1 \\ \end{array}\right) \label{M2}
\eeq
and define the reflection coefficient $\hat{\Gamma}=1-\hat{T}$. The infinite sum in eq. \ref{full_blast_q}
can be performed by working in the diagonal representation (denoted by $'$):
\beq
\vec{q}_{_{T}} = \frac{1}{2}\hat{M} \hat{T}' \hat{M}^{-1} \sum_{j=0}^{\infty} \hat{M} \hat{\Gamma}'^j \hat{M}^{-1}
\hat{M}\left(\begin{array}{c} \hat{M}^{-1}_{11}q
\\ \hat{M}^{-1}_{12}q \end{array}\right) \label{full_blast_qq}
\eeq
It is easy to show that:
\beq
\hat{\Gamma}'=\left(\begin{array}{cc} \Gamma_{11}-\Gamma_{12} & 0 \\ 0 &  \Gamma_{11}+\Gamma_{12}\\ \end{array}\right); ~~~ \hat{T}'= 1 - \hat{\Gamma}' \label{Gammas}
\eeq
Since $\hat{\Gamma}'$ is a diagonal matrix, it is very simple to take powers of it.
Expression \ref{full_blast_qq} can be thus written in a convenient form as:
\bear
\vec{q}_{_{T}} = && \frac{1}{2}\hat{M} \hat{T}' \hat{M}^{-1} \hat{M}
\left(\begin{array}{cc} \sum_{j=0}^{\infty}(\Gamma_{11}-\Gamma_{12})^j & 0 \\ 0 & \sum_{j=0}^{\infty}(\Gamma_{11}+\Gamma_{12})^j\\ \end{array}\right) \nonumber \\
&& \times \hat{M}^{-1} \hat{M}\left(\begin{array}{c} \hat{M}^{-1}_{11}q
\\ \hat{M}^{-1}_{12}q \end{array}\right) \label{full_blast_q2}
\eear
After performing the geometric sum and grouping terms:
\bear
\vec{q}_{_{T}}\!\! =\!\! \frac{1}{2}\hat{M} \! \Bigg[ \hat{T}' \!\!\! \left(\!\!\begin{array}{cc} \frac{1}{1\!\!-\!(\Gamma_{11}\!-\!\Gamma_{12})} & 0 \\ 0 & \frac{1}{1\!\!-\!(\Gamma_{11}\!+\!\Gamma_{12})} \\ \end{array}\!\!\right)\!\! \Bigg] \!\!\! \left(\!\!\begin{array}{c} \hat{M}^{-1}_{11}q
\\ \hat{M}^{-1}_{12}q \end{array}\!\!\right) \label{full_blast_q3}
\eear
where the term in brackets is the identity matrix. Thus:
\beq
\vec{q}_{_{T}} = \frac{1}{2}\hat{M} \left(\begin{array}{c} \hat{M}^{-1}_{11}
\\ \hat{M}^{-1}_{12} \end{array}\right) q \label{full_blast_q4}
\eeq
that finally yields:
\beq
\vec{q}_{_{T}} = \frac{1}{2} \left(\begin{array}{c} 1 \\ 0 \end{array}\right) q \label{full_blast_q4}
\eeq
as we wanted to demonstrate. The driven line carries away 1/2 of the signal per each strip end,
while the un-driven line does not carry net charge. The result requires of all reflections
to be considered.

\section{Current vs voltage generators in TLs}

\begin{figure}[ht!!!]
\centering
\includegraphics*[width=9.3 cm]{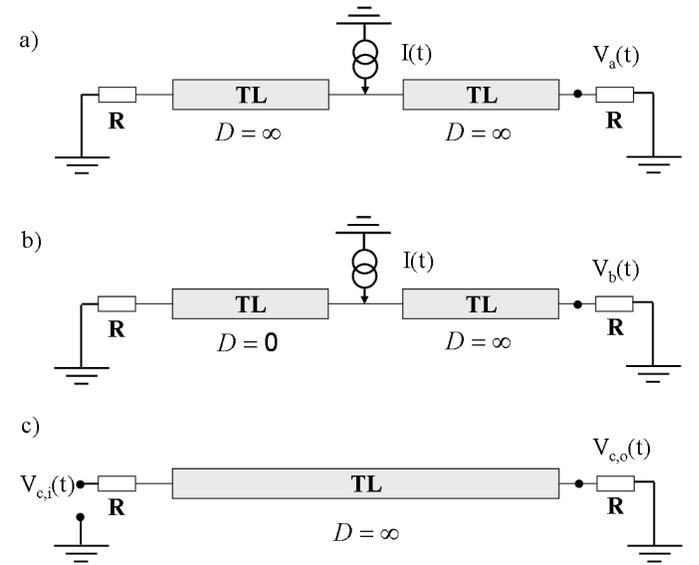}
\caption{The three cases discussed for illustrating the
equivalence between current and voltage sources.}
\label{app}
\end{figure}

Perhaps the clearest explanation of the relation between pulser
measurements and ideal current injection can be
obtained through Fig. \ref{app} for a 1-strip case. For the
sake of clarity, we assume that the line is loss-less
in the following reasoning.

Solutions to cases a) and b) can be directly obtained, without
solving eq. \ref{full_blast}. by properly combining the transmission 
and reflection coefficients:

\bear
& V_a(t) &=\frac{I(t)}{2}\times T_{i-o} \times R \\
& V_b(t) &=\frac{I(t)}{2}\times (1+(1-T_{i-o})) \times T_{i-o} \times R \nonumber \\
&& = \frac{I(t)}{2}\times T_{o-i} \times T_{i-o} \times R \label{b_case}
\eear
where:

\bear
&&T_{i-o} = \frac{2Z_c}{Z_c+R} \\
&&T_{o-i} = \frac{2R}{Z_c+R}
\eear
are the transmission coefficients from in-out and out-in of the transmission line (TL),
respectively. All terminating resistors are equal.
Under the typical situation $Z_c<R$, it is verified that $T_{i-o}<1$ and $T_{o-i}>1$. According
to the definitions in text:

\bear
&&F_{tr,a}(t) = \frac{1}{2}\frac{I(t)}{\tn{max}[I(t)]}\times T_{i-o} \times R \\
&&F_{tr,b}(t) = \frac{1}{2}\frac{I(t)}{\tn{max}[I(t)]}\times T_{o-i} \times T_{i-o} \times R
\eear
Perhaps counter-intuitively, transmission at the center of an un-matched loss-less 
strip (with $Z_c<R$) will be always smaller than at its ends
ends, when looking at the signal collected at the opposite end. The factor between the two
situations is $T_{o-i}$ and arises from the constructive interference between the direct and reflected
waves.

The solution in situation c) is also easy to find, being:

\beq
V_{c,o}(t) =\frac{V_{c,i}(t)}{2R}\times T_{o-i} \times T_{i-o} \times R \label{c_case}
\eeq
The voltage measured in absence of TL or with it being adapted is thus:

\beq
V(t)=V_{c,i}(t)/2
\eeq
as for a perfect voltage divider. The correspondence between
situations b) and c) (eq. \ref{b_case} and
eq. \ref{c_case}) can be obtained through the equivalence:

\beq
I(t) \equiv \frac{V_{c,i}(t)}{R}
\eeq
or:

\beq
I(t) \equiv 2 \frac{V(t)}{R} \label{ref_I2}
\eeq
that is eq. \ref{ref_I} in text. It gives the equivalence between the current
induced at one end of a TL, $I(t)$, and a pulser measurement, being $V(t)$
 the voltage drop measured in a matched load. Thus, the solutions for the former case
in eq. \ref{full_blast} can be directly applied for a pulser measurement after making the
equivalence \ref{ref_I2}.

\end{document}